\documentclass[longauth]{aa}   % for the long lists of affiliations

\usepackage{graphicx}
\usepackage{xcolor}
\usepackage{txfonts}
\usepackage{lipsum}
\usepackage{subcaption}
\usepackage{lscape}    
\usepackage{placeins}  
\usepackage[colorlinks=true,linkcolor=blue,citecolor=blue,urlcolor=blue]{hyperref}  

\renewcommand{\linenumbers}{} % disable any later calls
\renewcommand{\nolinenumbers}{} % avoid errors

\nolinenumbers

\begin{document}

\title{Time-resolved protoplanetary disk physics in DQ Tau with JWST}

\author{\'A. K\'osp\'al\inst{1,2,3}
\and
P. \'Abrah\'am\inst{1,2,4}
\and
V. V. Akimkin\inst{5}
\and
L. Chen\inst{1}
\and
J. Forbrich\inst{6}
\and
K. V. Getman\inst{7}
\and
B. Portilla-Revelo\inst{7, 8}
\and
D. Semenov\inst{9, 3}
\and
S. E. van Terwisga\inst{10}
\and
J. Varga\inst{1}
\and
L. Zwicky\inst{1,2}
\and
G. G. Bal\'azs\inst{1,2}
\and
Zs. Bora\inst{1}
\and
\'A. Horti-D\'avid\inst{1}
\and
A. P. Jo\'o\inst{1}
\and
W. Og\l{}oza\inst{11}
\and
B. Seli\inst{1}
\and
M. Siwak\inst{10}
\and
\'A. S\'odor\inst{1}
\and
N. Tak\'acs\inst{1}
}

\institute{Konkoly Observatory, HUN-REN Research Centre for Astronomy and Earth Sciences, MTA Centre of Excellence, Konkoly-Thege Mikl\'os \'ut 15-17, 1121 Budapest, Hungary\\
\email{kospal@konkoly.hu}
\and
Institute of Physics and Astronomy, ELTE E\"otv\"os Lor\'and University, P\'azm\'any P\'eter s\'et\'any 1/A, 1117 Budapest, Hungary
\and
Max-Planck-Insitut f\"ur Astronomie, K\"onigstuhl 17, 69117 Heidelberg, Germany
\and
Institute for Astronomy (IfA), University of Vienna,T\"urkenschanzstrasse 17, A-1180 Vienna, Austria
 \and
Institute of Astronomy, Russian Academy of Sciences, 48 Pyatnitskaya St., Moscow, 119017, Russia
\and
Centre for Astrophysics Research, University of Hertfordshire, College Lane, Hatfield, AL10 9AB, UK
\and
Department of Astronomy \& Astrophysics, Pennsylvania State University, 525 Davey Laboratory, University Park, PA 16802, USA
\and
Center for Exoplanets and Habitable Worlds, Penn State University, 525 Davey Laboratory, 251 Pollock Road, University Park, PA, 16802, USA
\and
Zentrum f\"ur Astronomie der Universit\"at Heidelberg, Institut f\"ur Theoretische Astrophysik, Albert-Ueberle-Str. 2, 69120 Heidelberg, Germany
\and
Space Research Institute, Austrian Academy of Sciences, Schmiedlstrasse 6, A-8042 Graz, Austria
\and
Mt. Suhora Astronomical Observatory, University of the National Education Commission, ul. Podchor\k{a}\.zych 2, 30-084 Krak\'ow, Poland
}

\date{Received 18 June 2025 / Accepted 26 August 2025}
 
\abstract
{}
{Accretion variability is ubiquitous in young stellar objects. While large ($2.5 - 6\,$mag) outbursts may strongly affect the disk structure, the effects of moderate ($1 - 2.5\,$mag) bursts are less understood. We aim to characterize the physical response of the disk around the  eccentric binary system DQ\,Tau to its periodic accretion changes.} 
{We organized a multi-wavelength observing campaign centered on four JWST/MIRI spectra. We targeted three consecutive periastrons (high accretion state) and one apastron (quiescence). We used optical and near-infrared spectroscopy and photometry to measure how the accretion luminosity varies. We decomposed the multi-epoch spectral energy distributions into stellar, accretion, and rim components. In the MIRI spectra, we fitted the solid-state features using various opacity curves and the molecular features using slab models.}
{We find the inner disk of DQ\,Tau to be highly dynamic. The temperature, luminosity, and location of the inner dust rim vary in response to the movement of stars and the $L_{\rm acc}$ variations ($0.10\,$L$_{\odot} - 0.40\,$L$_{\odot}$). This causes variable shadowing of the outer disk, leading to an anti-correlation between the rim temperature and the strength of the 10$\,\mu$m silicate feature. The dust mineralogy remains constant across all epochs, dominated by large ($>2\,\mu$m) amorphous olivine and pyroxene grains, with smaller fractions of crystalline forsterite. The excitation of CO ($1550-2260$\,K), HCN ($880-980\,$K), and hot H$_2$O ($740-860\,$K) molecules as well as the luminosity of the [NeII] line correlate with the accretion rate, while the warm ($\sim650\,$K) and cold ($\sim170-200\,$K) H$_2$O components are mostly constant. CO emission, originating from a hot ($>1500\,$K) region likely within the dust sublimation radius, is most sensitive to $L_{\rm acc}$ changes. In comparison with other T\,Tauri disks, DQ\,Tau is highly C-poor and displays moderately inefficient pebble drift.}
{We conclude that even moderate accretion rate changes affect the thermal structure in the planet-forming disk regions on short timescales, providing a crucial benchmark for understanding disk evolution.}


\maketitle

\section{Introduction}

Accretion variability is fundamentally important during the formation of Sun-like stars and their planetary systems. T\,Tauri stars were first recognized on the basis of their remarkable optical variability \citep{joy1945}, and since then it has become clear that this variability extends across many orders of magnitude in wavelength, duration, and power \citep{fischer2023}. The most extreme accretion variability of the outbursting EX\,Lupi-type and FU\,Orionis-type stars have been shown to significantly alter the structure and geometry of the disk and envelope \citep{mosoni2013, takami2018}, and the physical, mineralogical, and chemical properties of the planet-forming material in the disk where terrestrial planets may be forming \citep{visser2015, cieza2016, rab2017a, molyarova2018, lee2019, abraham2019, kospal2023, calahan2024, zwicky2024, cecil2024, laznevoi2025}.

In addition to the largest outbursts, accretion-related variability spans the parameter range down to timescales of a few hours and amplitudes of a few 0.1\,mag. Between 13-21\% of young disk-bearing stars show bursts likely associated with variations in the accretion flow and related hot shock regions on the stellar surface \citep{cody2022}. Stellar flares significantly overlap with the strength and timescale of the accretion fluctuations but originate from the stellar magnetosphere \citep{favata2005, getman2008a, getman2008b, getman2021}. While the sudden increases in heating and ionizing radiation are fascinating windows on the complex magnetohydrodynamic processes in the inner disk and the stellar surface \citep{rab2017b,washinoue2024}, their effects on the evolution of the star and the disk are much less studied than those of the large outbursts \citep{abraham2009,topchieva2025}. This is an important gap in our current knowledge.

Here, we present the first results of an optical, near-infrared, and mid-infrared spectroscopic and photometric monitoring study of DQ\,Tau, a system with predictable flares and accretion rate variations thanks to its binarity. We study the effects of moderate luminosity changes on the properties of the dust and gas content of the circumbinary protoplanetary disk.

\section{Our target: DQ\,Tau}

DQ\,Tau has, at first sight, a normal protoplanetary disk. Its dust mass is $M_{\rm dust} \sim 66\,M_{\oplus}$, with a characteristic radius of $R_c \sim 24-32\,$au \citep{ballering2019,long2019}. It is not very massive or large; at $d = 195$\,pc, it is relatively nearby. However, in a very small cavity in the inner disk ($R_{\rm in} = 0.13 - 0.35$\,au, \citealt{kospal2018, muzerolle2019}) lies an eccentric, extremely close binary of two nearly equal mass M0-type stars. The orbital period of the binary is $P = 15.80$\,days, the orbital eccentricity $e = 0.56$, while the total mass of the two stars is $M_{*,\rm tot} = 1.52\,M_{\odot}$ \citep{czekala2016,fiorellino2022}. The components' distance at periastron / apastron is 0.06 / 0.22\,au. At orbital phase $\phi$ = 0.85--1.05 (i.e., around periastron), suddenly but predictably, the system's X-ray, UV, optical, IR, and sub-mm continuum flux increases. The X-ray and radio flares are due to interactions between the young stars' magnetospheres \citep{salter2010, getman2011, getman2022, getman2023, pouilly2023}; the UV and optical brightenings are due to pulsed accretion \citep{gunther2002, muzerolle2019, fiorellino2022}; while the infrared brightening is due to heating of the inner dusty rim of the disk during the pulsed accretion events during periastrons \citep{kospal2018}. Magnetic reconnection-driven X-ray and UV photons play a crucial role in shaping the environments of young stars by ionizing, heating, and photoevaporating protoplanetary disks and primordial planetary atmospheres. X-rays, in particular, enhance disk ionization levels, potentially triggering magnetorotational instability (MRI), which governs gas turbulence, viscosity, accretion rates, and planetary migration. They also contribute to ion-molecular chemistry, dust grain sputtering, and the ionization of disk coronae --- processes that may help launch jets and disk winds. While many of these X-ray-driven effects are still largely theoretical \citep{glassgold2000,glassgold2007,alexander2014,owen2019,ercolano2021,woitke2024}, observational evidence supports some ionization-sensitive tracers, such as [Ne\,II] emission and variable HCO$^+$ abundances in a subset of disks \citep{cleeves2017,espaillat2023,waggoner2023}.

Hydrodynamical simulations show that in eccentric binaries with circumbinary disks, the accretion rate onto the components pulsates: disk matter is strongly perturbed near the apocenter of the binary, and the material falls within the Roche lobes of the stars about a half orbit later, near periastron \citep{artymowicz1994, artymowicz1996, gunther2002,sytov2011}. DQ\,Tau is the prototype of close eccentric pre-main sequence binaries with pulsed accretion. The accretion rate in DQ\,Tau close to periastron increases consistently as evidenced by various accretion tracers from UV and optical photometry to optical and near-infrared spectroscopy. For instance, based on VLT/XSHOOTER spectra, \citet{fiorellino2022} measured a quiescent accretion rate of $3\times10^{-9}\,$M$_{\odot}$/yr (accretion luminosity $L_{\rm acc} = 0.04$\,L$_{\odot}$) far from periastron and a peak accretion rate of $5\times10^{-8}\,$M$_{\odot}$/yr ($L_{\rm acc} = 0.73$\,L$_{\odot}$) close to periastron, a factor of 17 change. For context, this change is larger than the moderate amplitude burst of EX\,Lup in 2022 (accretion luminosity changed by a factor of 7), but smaller than its large outburst in 2008 (factor of 40, \citealt{cruzsaenzdemiera2023}).

\citet{kospal2018} monitored one of the periastron brightening events in DQ\,Tau simultaneously at optical (with Kepler/K2) and infrared (with Spitzer/IRAC) wavelengths. They found that during the accretion event, the luminosity of the central source went up by $L_{\rm acc} = 0.8\,$L$_{\odot}$. As a consequence, the inner dust disk experienced enhanced heating. Indeed, the authors found that based on blackbody fits to the Spitzer data, the temperature in the inner disk went up from 825\,K to 917\,K. Again, for comparison, the temperature at the inner edge of the dust disk in EX\,Lup went up from 900\,K to 1700\,K in the large outburst in 2008 \citep{abraham2009}. The increased disk heating in EX\,Lup led to the first direct observation of annealing of amorphous silicate dust grains to crystalline ones, detected as the appearance of the $10.0\,\mu$m and $11.3\,\mu$m features of crystalline forsterite \citep{abraham2009}. At the same time, the molecular emission lines also changed in the infrared spectrum of EX Lup: OH and H$_2$O became stronger during the outburst, while organics, such as C$_2$H$_2$ and HCN, disappeared, possibly due to UV photodissociation \citep{banzatti2012, banzatti2015,smith2025}.

EX\,Lup's outburst was more luminous and lasted longer than the periastron bursts of DQ\,Tau. But EX\,Lup-type outbursts are rare, while moderate luminosity changes due to fluctuating accretion in T Tauri-type stars are ubiquitous. We will use DQ\,Tau to study whether the smaller bursts may have similarly profound effects on the disk material as observed in EX\,Lup.

\section{Observations and Data Reduction}
\label{sec:data}

\subsection{JWST/MIRI spectroscopy}
\label{sec:miri}

\begin{figure*}[h!]
\centering
\includegraphics[width=\textwidth]{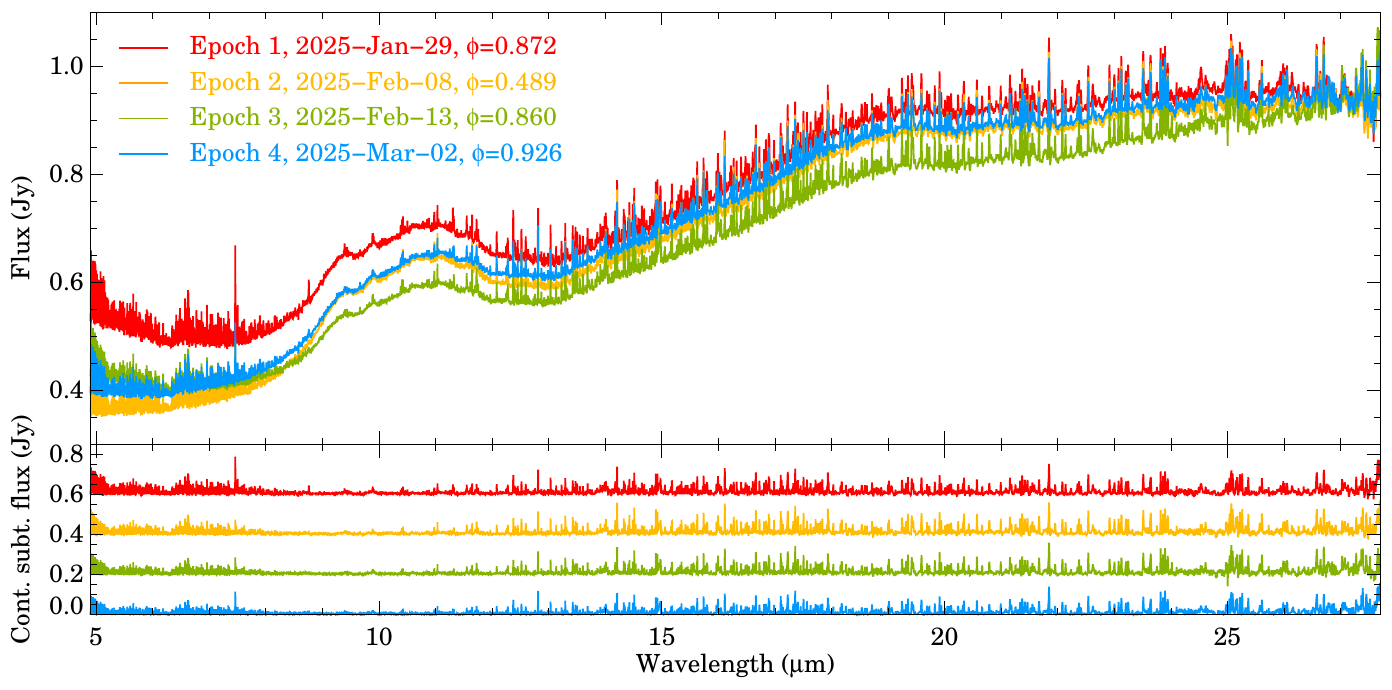}
\caption{JWST MIRI MRS spectra of DQ\,Tau. Various colors mark the different observing epochs, whose dates and the corresponding orbital phases are indicated in the upper left corner. The bottom panel shows the continuum-subtracted spectra shifted along the y axis for clarity.}\label{fig:comparison}
\end{figure*}

We conducted a 4-epoch mid-infrared spectroscopic monitoring of DQ\,Tau using the Mid-Infrared Instrument (MIRI, \citealt{miri}) of the James Webb Space Telescope (JWST, \citealt{jwst}) in the framework of the program GO 4876 (PI: \'A.~K\'osp\'al). The observations were time-constrained: epochs 1, 3, and 4 were scheduled in 3-day-long windows targeting three consecutive periastrons from orbital phase 0.85 to 1.01 to maximize chances of catching DQ\,Tau in a high accretion phase, while epoch 2 targeted an out-of-periastron date with low chance of any elevated accretion activity. We used three different grating settings to cover the whole 5--28\,$\mu$m wavelength range with the Medium-Resolution Spectrometer (MRS). We used target acquisition with the Neutral Density Filter to center our target precisely. The recommended 4-point dither pattern optimized for point sources was used for our science target. DQ\,Tau is too bright for the SLOWR1 read-out, therefore we used FASTR1. 

We reduced the JWST MIRI MRS spectra using a publicly available hybrid pipeline developed for the MINDS collaboration \citep{Christiaens2024}. The code is based on the standard JWST pipeline and uses algorithms from the Vortex Image Processing (VIP) package for background subtraction, bad pixel correction, and spectral extraction \citep{GomezGonzales2017,Christiaens2023}. Following the basic reduction, we subtracted a local background from the data cube at each wavelength, by defining a sky annulus around the centroid position of the source. We extracted 1D spectra assuming point source geometry and calculated aperture photometry with aperture sizes of 2 times the full width at half maximum. Then we corrected for flux losses outside the aperture using the default aperture corrections from the JWST pipeline, determined from high-resolution PSF models (10-15\%, Fig.~7 in \citealt{law2025}). Then we applied an additional residual fringe correction on the spectrum. We stitched together the neighboring bands  using the overlapping parts of the spectrum, taking the shortest band as reference. The scaling factors were first estimated by the pipeline, but we also checked them via visual inspection, and refined them at the $\leq$1\% level. We combined the 12 bands into a single spectrum by cutting the overlapping channels at the blue end of each band in most cases, but this was also interactively checked and adjusted when needed. We fitted a continuum to the spectrum using the method of \citet{Temmink2024} and subtracted it to produce a line spectrum. The final reduced MIRI spectra are displayed in Fig.~\ref{fig:comparison}.

\subsection{Ground-based optical and infrared spectroscopy}

We complemented the JWST observations with contemporaneous ground-based high-resolution optical spectroscopy using the Fiber-fed Extended Range Optical Spectrograph (FEROS) on the MPG 2.2\,m telescope located at ESO's La Silla Observatory (Chile), and with medium-resolution optical/near-infrared spectroscopy using SpeX on the NASA Infrared Telescope Facility (IRTF, Mauna Kea, USA). Our FEROS data were collected in the framework of the program 0114.C-6020(A) or 114.28H7.001 (PI: D.~Semenov) and our SpeX data in 2024B051 and 2025A019 (PI: \'A.~K\'osp\'al).

FEROS is a high-resolution \'echelle spectrograph that covers the optical ($360-920$\,nm) wavelength range at $R=48\,000$. In one observing block, we took three exposures with 500\,s each using the object-sky mode, when one fiber recorded signal from DQ\,Tau and one from a nearby sky background to enable subtraction of sky emission lines. The Atmospheric Dispersion Corrector (ADC) was used to correct for the effect of atmospheric dispersion in this case. We also took three exposures with 500\,s each using the object-cal mode, when one fiber recorded signal from DQ\,Tau and one the signal of a Thorium-Argon calibration lamp. We executed one or two such observing blocks each night between 2025 January 28-30, February 8-10, February 14-16, and March 1-3. We downloaded the calibrated spectra from the ESO archive processed with the FEROS Data Reduction System (DRS), implemented within ESO-MIDAS. The spectrum taken on 2025 March 1 is faulty, therefore, we exclude it from further analysis. 

IRTF/SpeX \citep{rayner2003} observations were scheduled over 30 nights between 2025 January 20 and March 2, out of which data could be taken on 21 nights. At all epochs, we used the SXD grism with the $0\farcs5\times15''$ slit, providing $R=1200$ resolution in the $0.7-2.5\,\mu$m wavelength range. We used an exposure time of 60\,s and took 10-24 exposures using the usual ABBA nodding pattern along the slit to enable sky subtraction. On two of the nights, 2025 February 5 and 8, we also used the LXD\_long grism with the $0\farcs5\times15''$ slit, providing $R=1500$ resolution in the $2.0-5.3\,\mu$m wavelength range. In this setup, we used an exposure time of 2\,s, 10-12 coadds, and took 16-20 exposures. We used either k\,Tau or 18\,Ori (both A0-type stars) as telluric calibrator, observed with the same instrumental setup as the science observations. We reduced the spectra using Spextool \citep{cushing2004,vacca2003}, consisting of steps for flatfield correction and wavelength calibration, sky subtraction, extraction of the 1D spectra, telluric correction, and combination of the \'echelle orders into one continuous spectrum.

\subsection{Optical and near-infrared photometry}

We complemented the JWST/MIRI observations with contemporaneous optical/near-infrared photometry from Swift's Ultraviolet/Optical Telescope (UVOT, proposal ID 26200161, PI: K.~Getman), from the Piszk\'estet\H{o} Mountain Station of Konkoly Observatory (Hungary), Mount Suhora Observatory of the Cracow Pedagogical University (Poland), the Planetarium and Astronomical Observatory of the Silesian Science Park in Chorz\'ow (Poland), the Rapid Eye Mount (REM, proposal ID 24B021, PI: \'A.~K\'osp\'al) telescope at ESO's La Silla Observatory (Chile), as well as with data collected by (amateur) astronomers across the world through the American Association of Variable Star Observers (AAVSO).

We conducted multiple snapshot observations of DQ\,Tau during three periastron passages of interest using the Neil Gehrels Swift Observatory \citep{gehrels2004}. The Swift Ultraviolet/Optical Telescope (UVOT) is a 30\,cm Ritchey-Chretien telescope that provides ultraviolet and optical coverage in a $17'\times17'$ field-of-view simultaneously with Swift observations. UVOT was operated in the 0x30ed standard six-filter, blue-weighted mode with a 6-hour cadence, obtaining 12 data points each between 2025 January 28-31, February 13-16, and March 1-4. UVOT magnitudes in the six filters ($V$, $B$, $U$, $W1$, $M2$, $W2$) were derived using the uvotsource tool from HEASoft (v6.33.2), together with the Swift UVOT CALDB (version 20240201).

Our optical ($Bg'Vr'i'$ filters) monitoring from Piszk\'estet\H{o} Observatory covered 23 clear nights from 2025 January 17 until March 3 using the Astrosysteme ASA800 80\,cm RC telescope (RC80) equipped with an FLI Microline CCD camera (pixel scale: $0\farcs55$/pixel, field of view: $18\farcm4\times18\farcm7$). Observations in Poland used the 60\,cm Carl-Zeiss telescope with an Apogee Aspen-47 camera (pixel scale: $1\farcs117$/pixel, field of view: $19\farcm1\times19\farcm1$) and the 70\,cm PlaneWave CDK 700\,mm telescope with an FLI Kepler 4040 camera (pixel scale: $1\farcs178$/pixel, field of view: $40\farcm2\times40\farcm2$). Data were taken on the nights of 2025 January 22, 27, 28, 29, February 8, 9, 10, and March 2 with $BVR_CI_C$ filters. We monitored DQ\,Tau in the optical $g'r'i'z'$ and near-infrared $JHK$ filters using the Rapid Eye Mount (REM) telescope on 40 clear nights from 2025 January 17 until March 9. REM is a 60\,cm telescope equipped with ROS2, a multi-channel imaging camera (pixel scale: $0\farcs58$pixel, field of view: $9\farcm9{\times}9\farcm9$) and REMIR, a Hawaii I near-infrared camera (pixel scale: $1\farcs235$/pixel, field of view: $10'\times10'$). On 13 of the nights, only near-infrared data were taken.

With the RC80, REM/ROS2, and Polish telescopes, at each observing night, 3-9 images were taken with each filter in case of nightly snapshots and up to 300 images in case of high-cadence monitoring, with exposure times between 1--90\,s. The data was reduced following the usual steps for bias, dark, and flatfield correction. Aperture photometry was extracted in a 7 pixel radius aperture for RC80 and REM and 5 pixel radius for the Polish telescopes, with sky annulus between 20 and 40 pixels. Instrumental magnitudes were calculated this way for for DQ\,Tau and 5-10 comparison stars in the field of view. The APASS9 magnitudes \citep{henden2015} of the comparison stars were used for the photometric calibration of the instrumental magnitudes of DQ\,Tau. There was generally a good agreement between photometry from RC80 and REM. After checking data from different telescopes obtained on the same nights, we applied shifts of +0.12\,mag in $B$ and +0.02\,mag in $V$ for the Polish telescopes and $-$0.10\,mag in $V$ for UVOT to match them to the RC80 and REM magnitudes.

During the REM/REMIR observations, $JHK$-band images were taken with a 5-point dither pattern to allow for sky subtraction. Exposure times were 2\,s in $J$ and $0.8-2$\,s in $HK$. Data reduction steps included sky subtraction, registration, and coadding exposures by dither position and filter. For photometric calibration, we used the Two Micron All Sky Survey (2MASS) catalog \citep{cutri2003}. We calculated instrumental magnitudes for DQ Tau and three good-quality 2MASS stars in the field-of-view using an aperture radius of 3 pixels and a sky annulus between 9 and 12 pixels.

\begin{figure*}[h!]
\centering
\includegraphics[width=\textwidth]{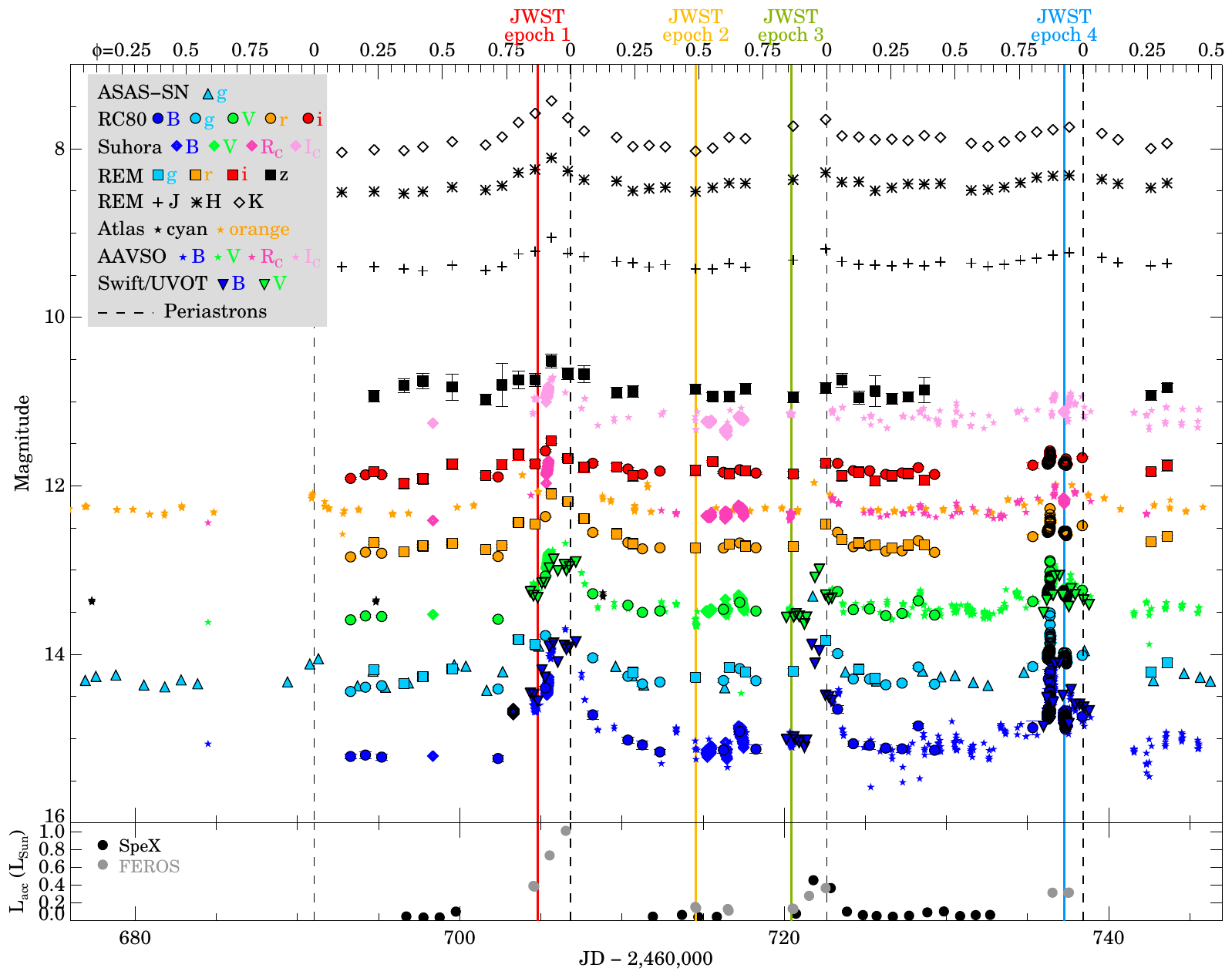}
\caption{Light curves (top) and accretion luminosity (bottom) of DQ\,Tau. Vertical dashed lines mark the epochs of periastrons. The solid vertical lines in color show the epochs of the JWST/MIRI observations.}\label{fig:light}
\end{figure*}

\section{Results and Analysis}

\subsection{Brightness and accretion rate evolution}
\label{sec:acc}

\begin{figure*}[h!]
\centering
\includegraphics[width=\textwidth]{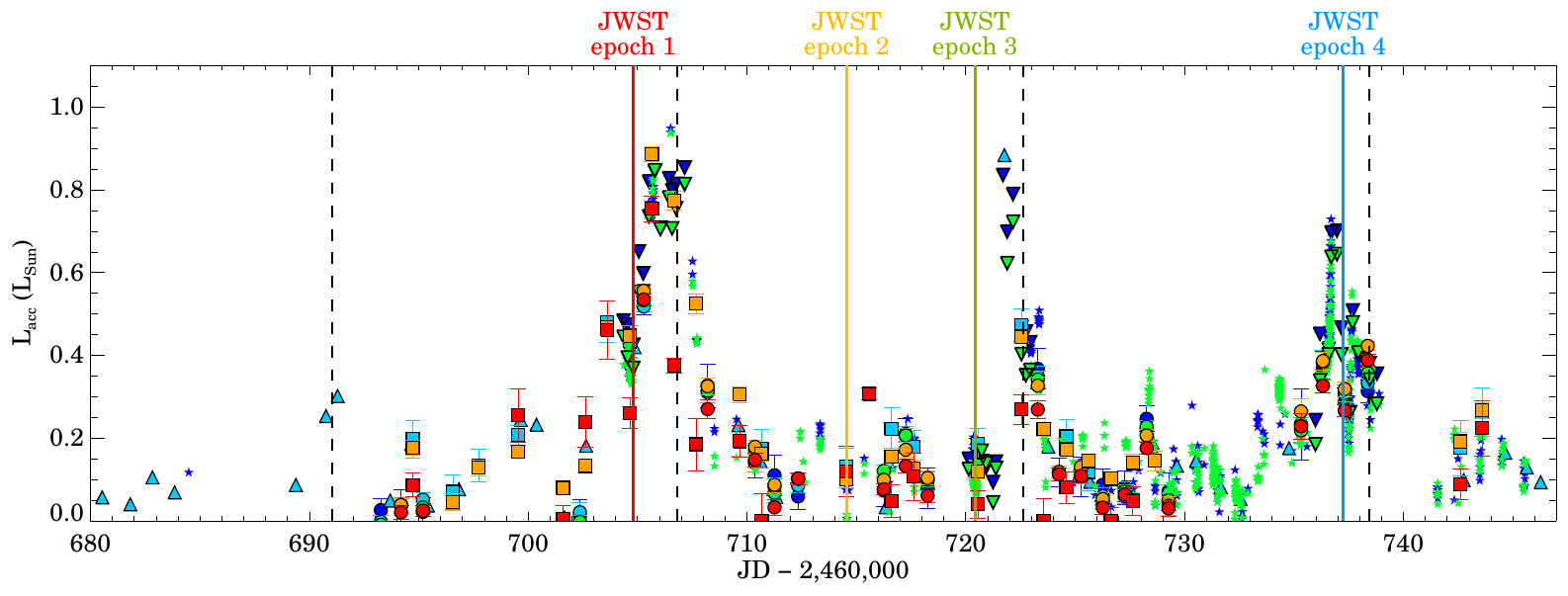}
\caption{Accretion luminosity of DQ\,Tau as a function of time. The $g$ (cyan), $V$ (green), $r$ (orange), and $i$ (red) data points were first scaled to match the $B$ (blue) filter, then scaled again to convert the B magnitudes to $L_{\rm acc}$. Symbol types are as in Fig.~\ref{fig:light}. For more details, see Sec.~\ref{sec:acc}.} \label{fig:lacc}
\end{figure*}

The results of our photometric monitoring covering three consecutive periastrons is presented in Fig.~\ref{fig:light}. Using the orbital parameters from \citet{czekala2016}, the periastrons in this time range occurred on 2025 January 31 at 08:01:14 UT (JD = 2460706.83), on 2025 February 16 at 03:15:31 UT (JD = 2460722.64), and on 2025 March 3 at 22:29:48 UT (JD = 2460738.44). These times are marked by vertical dashed lines in Fig.~\ref{fig:light}. As expected, DQ\,Tau produced brightening events close to each periastron, with very different light curve shapes during the individual events. Leading up to the first periastron, there was a slow brightening with a smaller dip on the light curves right around the time of the epoch 1 JWST observation. Peak brightness occurred close to periastron, followed by a slow fading. During the epoch 2 JWST measurement, DQ\,Tau was in quiescence. Before the next periastron, DQ\,Tau brightened rather quickly, reaching a peak before the periastron at around $\phi=0.9$, followed by similarly steep fading. This event was so short that our epoch 3 JWST observation caught DQ\,Tau still in quiescence, where its optical photometry was close to those around JWST epoch 2.

DQ\,Tau displayed a rather complex behavior around the last periastron. There was a brief brightening peaking at 2460736.39 displaying classical flare-like characteristics, such as a linear rise followed by an exponential decay \citep{hawley2014,kospal2018}. The brightening in the $B,g',V,r',i'$ filters were 0.64, 0.50, 0.36, 0.26, and 0.14 mag, or a factor of 1.81, 1.58, 1.39, 1.27, and 1.13, respectively. This flare seems to be superimposed onto a slower brightening, as data between 2460736.5 and 2460736.7 shows. The next data points from 2460737.2 (simultaneously with the epoch 4 observations of JWST) again show lower brightness and a slow fading, suggesting that the peak of this event was some time between 2460736.7 and 2460737.0, when we have no data. Brightness levels during JWST epoch 4 were still above the typical quiescent brightness, but lower than the peaks we measured during our monitoring.

To estimate the accretion rate and its variations over our monitoring program, we analyzed our IRTF/SpeX and MPG\,2.2\,m/FEROS spectra. For SpeX, we applied the same method as described in \citet{muzerolle2019}. We first fitted simultaneously the veiling and the $A_V$ difference compared to a non-accreting template spectrum, for which we used a SpeX observation of LkCa\,21 ($A_V = 0.3$\,mag), the same star as was used in \citet{muzerolle2019}. We then subtracted the veiled template from the spectrum of DQ\,Tau and corrected the spectrum for $A_V$ using the reddening curve from \citet{cardelli1989}. Finally, we measured the fluxes of four lines, Paschen $\beta$, $\gamma$, $\delta$, and Brackett $\gamma$, converted the line fluxes to line luminosities as $L_{\rm line} = 4{\pi}d^2F_{\rm line}$, and converted the line luminosities to accretion luminosities using the scaling relationships from \citet{alcala2017}. The final accretion luminosities, obtained as the average of accretion luminosities from the four emission lines we used, are plotted in the bottom panel of Fig.~\ref{fig:light} and listed in Tab.~\ref{tab:irtf}. The obtained $A_V$ values ranged from 1.04\,mag to 2.15\,mag, with an average of $1.42\,$mag. Considering the typical uncertainty of our $A_V$ estimates of 0.18\,mag, the variations in $A_V$ are not significant, remaining below 2$\sigma$. Quiescent accretion luminosities are between 0.03 and 0.11\,L$_{\odot}$, while two SpeX spectra, those on 2025 February 15-16, caught the object during a periastron brightening and resulted in accretion luminosities of $0.40-0.48\,$L$_{\odot}$.

In the wavelength range of FEROS, the Balmer H$\alpha$ and the CaII IRT lines could be used to estimate accretion rates. The CaII IRT were not always detectable in our FEROS spectra, therefore we only use the H$\alpha$ line. These line profiles are displayed in Fig.~\ref{fig:halpha}. We followed the same procedure as before for the SpeX spectra: we first measured the line flux, then converted it to line luminosity, then to accretion luminosity. As the FEROS spectra do not allow an independent estimate of $A_V$ and as we already established that $A_V$ is constant, we corrected the H$\alpha$ line fluxes for a fixed $A_V=1.42$\,mag. The resulting accretion luminosities are plotted in Fig.~\ref{fig:light} and listed in Tab.~\ref{tab:feros}. Based on FEROS, accretion luminosities ranged between 0.11 and 1.01\,L$_{\odot}$. Considering the different accretion tracers, different spectral resolutions, different wavelength ranges, and different ways of absolute flux calibration of SpeX and FEROS, there is generally a very good agreement between the two sets of accretion luminosities for the common epochs, as can be seen in the bottom panel of Fig.~\ref{fig:light}.

As the photometric brightness variations indicate, accretion rate may change on much shorter time scales than the cadence of our spectroscopic observations. Therefore, we establish a relationship between the accretion luminosity and the $B$-band light curve, the shortest wavelength with good temporal coverage. First, we made scatter plots using the Konkoly/RC80 observations for $B$ vs.~$g'$, $B$ vs.~$V$, $B$ vs.~$r'$, and $B$ vs.~$i'$. The distribution of data points in these plots could be well fitted  with the following linear relationships: $B = -3.35(0.09) + 1.29(0.03) \times g'$, $B = -8.07(0.16) + 1.72(0.04) \times V$, $B = -10.51(0.25) + 2.01(0.07) \times r'$, $B = -20.11(0.60) + 2.98(0.16) \times i'$ (the uncertainties of the slopes and intercepts are given in parentheses). These equations allowed us to convert the $g'$, $V$, $r'$, and $i'$ band light curves to the $B$ filter. Based on this data set, we interpolated $B$ band magnitudes for the epochs of the FEROS spectra and fitted the following relationship between the $B$ magnitude and the accretion luminosity: $L_{\rm acc} / L_{\odot} = 9.28 - 0.61 \times B$. This allowed us to convert the light curve to an accretion luminosity curve. This is presented in Fig.~\ref{fig:lacc}. We then used this figure to estimate the accretion luminosities at the time of our JWST observations and obtained the following values: $L_{\rm acc,1}$ = 0.40\,L$_{\odot}$, $L_{\rm acc,2}$ = 0.10\,L$_{\odot}$, $L_{\rm acc,3}$ = 0.15\,L$_{\odot}$, and $L_{\rm acc,4}$ = 0.35\,L$_{\odot}$ in the four JWST epochs, respectively, with uncertainty of about 0.03\,L$_{\odot}$. As expected, epoch 2, obtained close to apastron, has the lowest accretion luminosity, while in the other three epochs we detect various levels of increased accretion, as expected close to periastron. In Sec.~\ref{sec:gaslines}, we will derive accretion luminosities from the JWST spectra themselves, using the H\,I $10-7$ line.

\subsection{Variability of the spectral energy distribution}
\label{sec:sed}

\begin{figure*}
\centering
\includegraphics[width=0.49\textwidth,page=1]{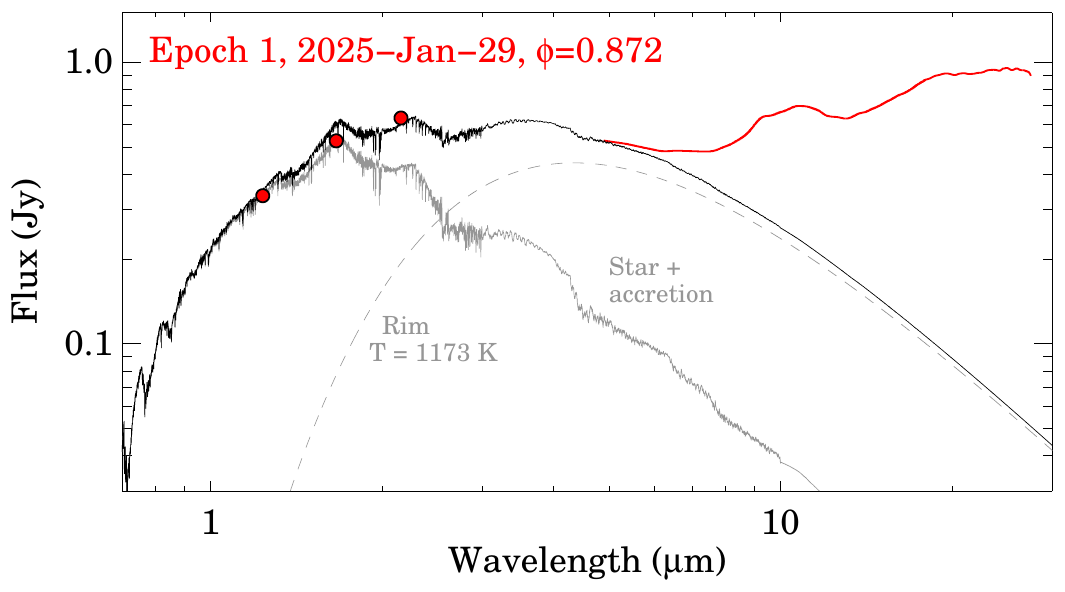}
\includegraphics[width=0.49\textwidth,page=2]{sed_per_epoch.pdf}
\includegraphics[width=0.49\textwidth,page=3]{sed_per_epoch.pdf}
\includegraphics[width=0.49\textwidth,page=4]{sed_per_epoch.pdf}
\caption{Spectral energy distributions of DQ\,Tau at the four epochs of the JWST/MIRI observations. The spectra between $0.7-5.3\,\mu$m are from IRTF/SpeX, while the spectra between $4.9 - 27.5\,\mu$m are the line-free MIRI spectra. Colored dots are REM $JHK$ and SpeX $L'$ photometric points. The solid gray spectra are models constructed to account for the stellar and accretion contributions. The dashed gray curves are blackbodies with the indicated temperatures, representing emission from the inner rim of the dusty disk. Solid black curves are the sum of the stellar, accretion, and rim contributions.}\label{fig:sed}
\end{figure*}

The broad-band spectral energy distribution (SED) of DQ\,Tau displayed significant variability over our monitoring. Fig.~\ref{fig:comparison} shows that the MIRI spectra intersect each other, suggesting multiple pivot points, i.e., wavelengths below/above which the SED brightened/faded or vice versa. For instance, the MIRI spectrum in epoch 2 (yellow curve) is the lowest at $\lambda < 8.5\,\mu$m, but the spectrum in epoch 3 (green curve) is lowest at $\lambda > 8.5\,\mu$m. Infrared pivot points are often explained by variable scale height of the inner rim of the dust disk, which casts a variable shadow to the outer disk \citep{flaherty2010, kospal2012}. To investigate this possibility, we constructed multi-epoch SEDs of DQ\,Tau for the JWST epochs covering the optical or near-infrared to the mid-infrared regime. In all epochs, we had $JHK$ photometry from REM contemporaneously with the MIRI spectra. In addition, in epoch 2, we also had $L'$ photometry from the SpeX guide camera, as well as SpeX SXD and LXD\_long spectra. In epoch 3, we had $L'$ photometry from the SpeX guide camera and a SpeX SXD spectrum. These SEDs are shown in Fig.~\ref{fig:sed}. In this figure and in the following analysis, we use the line-free continuum MIRI spectra derived in Sec.~\ref{sec:miri}.

To understand the various emitting components contributing to the total SED of DQ\,Tau, we made an attempt to decompose the SEDs into stellar, accretion, and rim components. In Sec.~\ref{sec:acc}, we used the SpeX observation of LkCa\,21 as a template to calculate veiling and reddening for each measured SpeX spectrum of DQ\,Tau. This procedure provides a veiled template spectrum that reproduces the spectrum of DQ\,Tau in epoch 2 (the quiescent apastron epoch) fairly well at $\lambda < 2\,\mu$m. We extrapolated this spectrum for $\lambda < 0.69\,\mu$m and $\lambda > 2.55\,\mu$m with an appropriately scaled NextGen model using $T_{\rm eff} = 3500\,$K, ${\log}g = 4$, and solar metallicity \citep{hauschildt1999}. The total unreddened luminosity of this model is $L_{\rm tot,2} = 1.42\,$L$_{\odot}$, out of which $L_{\rm acc,2} = 0.10\,$L$_{\odot}$ should be the accretion luminosity, as we determined in Sec.~\ref{sec:acc}. For Epochs 1, 3, and 4, we constructed similar veiled templates by scaling them so that their total luminosities are $L_{\rm tot,1} = 1.72\,$L$_{\odot}$, $L_{\rm tot,3} = 1.47\,$L$_{\odot}$, and $L_{\rm tot,4} = 1.67\,$L$_{\odot}$. This assumes that these spectra represent the sum of the luminosity of the two stars (assumed to be constant) and the accretion luminosity (which varies as we determined in Sec.~\ref{sec:acc}). These template spectra (marked as ``Star+accretion'') are plotted with solid gray curves in Fig.~\ref{fig:sed}.

We subtracted the template spectra from the observed  $JHKL'$ fluxes, SpeX SXD and LXD\_long spectra (where available) and the 4.9$\,\mu$m flux as measured by the shortest wavelength pixels of JWST/MIRI. We fitted the residuals with a single temperature blackbody, assumed to represent the temperature of the hottest dust grains in the system at that moment, i.e., the temperature of the inner dust rim. We obtained $T_{\rm rim,1} = 1173\pm7\,$K, $T_{\rm rim,2} = 996\pm9\,$K, $T_{\rm rim,3} = 1237\pm11\,$K, $T_{\rm rim,4} = 1097\pm8\,$K. The corresponding blackbody emission is plotted with dashed gray curves in Fig.~\ref{fig:sed}. The dereddened luminosity of this rim component is $L_{\rm rim,1} = 0.64\,$L$_{\odot}$, $L_{\rm rim,2} = 0.31\,$L$_{\odot}$, $L_{\rm rim,3} = 0.52\,$L$_{\odot}$, $L_{\rm rim,4} = 0.40\,$L$_{\odot}$. These results suggest that the rim was coldest and least luminous at apastron. This rim component we identify in DQ\,Tau is analogous to the near-infrared bump first identified in Herbig Ae/Be stars \citep{natta2001} then in T Tauri stars as well \citep{muzerolle2003}: an excess emission peaking at about $3\,\mu$m, well reproduced by a single temperature blackbody with $T=1500\,$K, identified as emission from the curved inner edge of the dust disk. The dust rim component in DQ\,Tau peaks at slightly longer wavelength and have a lower temperature because its inner edge is probably shaped by dynamical effects from the central eccentric binary rather than radiation only.

Fig.~\ref{fig:sed} shows that DQ\,Tau has excess emission over the stellar, accretion, and rim contributions at $\lambda > 4.9\,\mu$m, suggesting further emission components in the system. This could be the cold, optically thick disk midplane and the optically thin disk atmosphere, which also give rise to the silicate emission feature most readily observable around 10$\,\mu$m. In the following subsection, we study the variability and mineralogical composition of this feature and other solid-state features observable in the MIRI wavelength range.

\subsection{Dust mineralogy}

\begin{figure}
\centering
\includegraphics[width=\columnwidth]{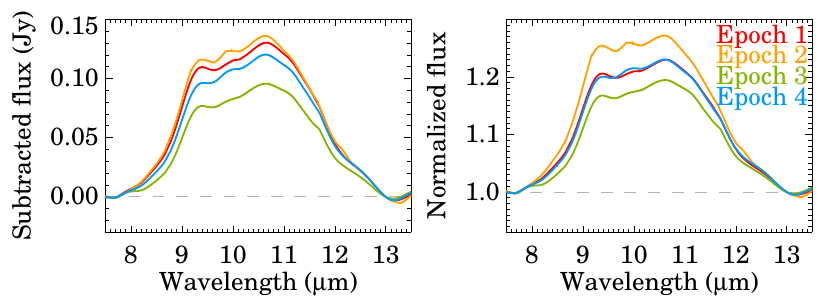}
\caption{Variability of the strength of the 10$\,\mu$m silicate feature in DQ\,Tau. Linear continua were subtracted (left) or were used to normalize the spectra (right).}\label{fig:sifeature}
\end{figure}

To study the variability of the 10$\,\mu$m silicate feature in DQ\,Tau, we calculated continuum-subtracted and continuum-normalized spectra using the line-free MIRI spectra derived in Sec.~\ref{sec:miri}. These spectra are devoid of atomic and molecular emission lines but still contain such broad spectral features as the 10$\,\mu$m silicate feature. We used the $7.5-7.9\,\mu$m and $12.9-13.1\,\mu$m ranges to fit linear continua. Results with second-order polynomial continua are similar. Fig.~\ref{fig:sifeature} suggests that the 10$\,\mu$m silicate feature was strongest at apastron, in epoch 2, while it was weakest in epoch 3, both in absolute flux and with respect to the local continuum. While the strength of the feature varies with time, its shape is remarkably constant, suggesting a non-variable dust composition with variable illumination/shadowing from the central source (see the discussion in Sec.~\ref{sec:discussion}). Nevertheless, in the following, we make a mineralogical decomposition in each epoch separately.

We fitted the solid-state features in the MIRI spectra of DQ\,Tau using two separate sets of opacity curves, differing in the choice of optical data. The opacities were calculated with optool \citep{Dominik2021}, using a distribution of hollow spheres (DHS) grain shape model \citep{Min2005}. The first opacity set, termed oldDHS, includes the following dust types: amorphous Mg$_2$SiO$_4$ (olivine-like, \citealt{Jager2003}), amorphous MgSiO$_3$ (pyroxene-like, \citealt{Jager1998}), crystalline forsterite \citep{Suto2006}, crystalline enstatite \citep{Dorschner1995}, and amorphous silica \citep{Kitamura2007}. For the second set, termed newDHS, we use optical measurements of forsterite and enstatite from a more recent paper \citep{Zeidler2015}, while the other opacities are the same as in the oldDHS set. Each set had three different grain sizes: $0.1\ \mu$m, $2.0\ \mu$m, and $5.0\ \mu$m. Therefore, in each set, we fitted $5\times3=15$ different opacity curves. The results of the fits are summarized in the Appendix in Figs.~\ref{fig:oldDHS}-\ref{fig:newDHS} and Tables \ref{tab:oldDHS}-\ref{tab:newDHS}. Both sets of models provide good fits with residuals below $2\%$. The reduced $\chi^2$ values are a bit lower for the newDHS set ($8.88-12.03$ vs.~$11.35-16.81$ for the oldDHS set). The resulting mass fractions are consistent across the epochs; there was no significant mineralogical variability during our monitoring. The newDHS and oldDHS results are not perfectly consistent, but there is an overall good agreement. We firmly detect the following dust components: $55-75\%$ $2-5\ \mu$m-sized amorphous Mg$_2$SiO$_4$ (olivine); $9-14\%$ $2\ \mu$m-sized amorphous MgSiO$_3$ (pyroxene); and $5-20\%$ large ($5\ \mu$m) crystalline enstatite.

\subsection{Gas lines and their variability}
\label{sec:gaslines}

The JWST/MIRI spectra of DQ\,Tau are very line-rich. Atomic H\,I lines can be found throughout the spectrum; in increasing wavelength, we detect the H\,I $10-6$ (Humphreys $\delta$) line at 5.129$\,\mu$m, H\,I $9-6$ (Humphreys $\gamma$) at 5.908$\,\mu$m, H\,I $14-7$ at $5.957\,\mu$m, H\,I $13-7$ at $6.292\,\mu$m, H\,I $12-7$ at $6.772\,\mu$m, H\,I $6-5$ (Pfund $\alpha$) at $7.460\,\mu$m, H\,I $8-6$ (Humphreys $\beta$) at $7.503\,\mu$m, H\,I $11-7$ at $7.508\,\mu$m, H\,I $10-7$ at $8.760\,\mu$m, H\,I $13-8$ at $9.392\,\mu$m, H\,I $12-8$ at $10.503\,\mu$m, H\,I $9-7$ at $11.309\,\mu$m, H\,I $7-6$ (Humphreys $\alpha$) at $12.372\,\mu$m, H\,I $11-8$ at $12.387\,\mu$m, H\,I $10-8$ at $16.209\,\mu$m, H\,I $8-7$ at $19.062\,\mu$m, and H\,I $11-9$ at $22.340\,\mu$m. Some of these are marked in Fig.~\ref{fig:h_comparison} and in Figs.~\ref{fig:co}-\ref{fig:h2o} in the Appendix. Further atomic lines we detected are the forbidden [Ar\,II] line at $6.985\,\mu$m and [Ne\,II] at $12.814\,\mu$m (Fig.~\ref{fig:forbidden}). We searched for the [S\,IV] line at $10.511\,\mu$m, [Ne\,III] at $15.555\,\mu$m, [S\,III] at $18.713\,\mu$m, and [S\,I] at $25.249\,\mu$m, but these lines are undetected. H$_2$ lines are most likely not detected. We checked the portions of the spectra around the wavelengths of the S(1) to S(8) lines, but if any feature was present, those could be readily explained by emission from H$_2$O, OH, or CO.

The $\lambda < 5.3\,\mu$m range is dominated by the P-branch of the fundamental lines of CO with minor contributions from H$_2$O (Fig.~\ref{fig:co_comparison}). In the $5.3 - 9.0\,\mu$m range, most lines are rovibrational lines of H$_2$O. The $9.0 - 12.0\,\mu$m range of the spectra suffer from residual detector fringing, but further H$_2$O lines (both rovibrational and pure rotational), as well as rotational lines of hot OH are clearly identifiable. From $12\,\mu$m, we find mostly pure rotational H$_2$O lines and the lines of OH, but the R-branch lines of HCN also start appearing. The strongest feature of HCN is the Q-branch peak of the $v_2$ bending mode at $14.0\,\mu$m \citep{bruderer2015}. A similar Q-branch $v_2$ feature from CO$_2$ can be seen at $15.0\,\mu$m \citep{bosman2017}. The Q-branch $v_5$ feature from C$_2$H$_2$ at $13.7\,\mu$m is not detected. The rest of the spectrum, above $\lambda > 15\,\mu$m, is dominated by H$_2$O and OH lines. We searched for features from additional molecules such as CH$_3^+$, CH$_4$, C$_2$H$_4$, C$_2$H$_6$, C$_4$H$_2$, HC$_3$N, NH$_3$, and HCO$^+$, but found no detection of them.

\begin{figure}
\centering
\includegraphics[width=0.47\textwidth,page=3]{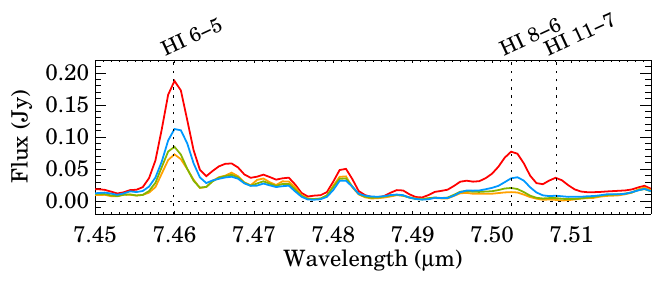}
\caption{ Examples for atomic H lines in the spectra of DQ\,Tau. Colors encode the different epochs as in Fig.~\ref{fig:sifeature}.}\label{fig:h_comparison}
\end{figure}

The H\,I lines display significant variability (Fig.~\ref{fig:h_comparison}), the brighter lines are clearly strongest in epoch 1, followed by epoch 4, then epoch 3, and weakest in epoch 2, in correlation with the variations of the accretion rate (Sec.~\ref{sec:acc}). The weaker lines are only detected in epochs 1 and 3. In these epochs, we measured the luminosity of the H\,I $10-7$ line at $8.760\,\mu$m and converted them to accretion luminosity using the empirical relation derived by \citet{tofflemire2025} for DQ\,Tau. The resulting luminosities are 0.41\,L$_{\odot}$ and 0.09\,L$_{\odot}$ for epochs 1 and 4, respectively. The former one is in good agreement with the accretion we determined for epoch 1 above, while the latter one is more discrepant, perhaps due to the low signal-to-noise of this line.

\begin{figure}
\centering
\includegraphics[width=0.49\textwidth,page=1]{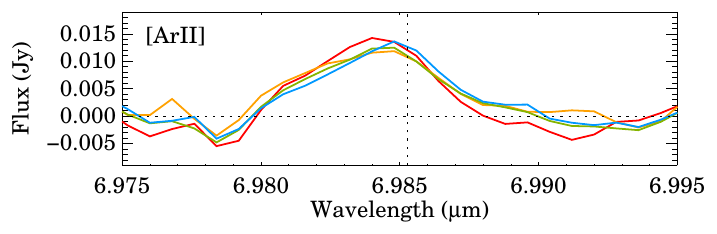}\\
\includegraphics[width=0.45\textwidth,page=2]{forbiddenlines.pdf}
\caption{Forbidden lines in the spectra of DQ\,Tau. For the [ArII] line, we first subtracted the best-fitting H$_2$O slab model from the spectra, while the [NeII] line is free from H$_2$O contamination.}\label{fig:forbidden}
\end{figure}

The forbidden [Ar\,II] and [Ne\,II] lines show a hint for variability (Fig.~\ref{fig:forbidden}). Fitting Gaussians to the line profiles and measuring their peak location, we found that the lines are blueshifted by $-33$ to $-77$\,km\,s$^{-1}$, suggesting some wind contribution. The luminosities of the [Ne\,II] line are $\log_{10}L_{[\rm NeII]}/L_{\odot} = -4.735\pm0.017, -4.797\pm0.019, -4.789\pm0.018, -4.769\pm0.018$. This suggests that the line varied little: there is a $\sim3\sigma$ hint that the [Ne\,II] line was stronger in epoch 1 than in epochs 2-3. The obtained luminosities fit well into the trend of [Ne\,II] vs.~accretion luminosities \citet{xie2023} collected for late M-type stars and earlier type stars, suggesting that the core of the line traces the disk surface ionized by the EUV and/or X-ray irradiation from the central binary rather than jets or winds. This is also supported by the fact that the [Ne\,II] line luminosity shows a tight correlation with the accretion luminosity in DQ\,Tau (see the discussion in Sec.~\ref{sec:discussion}).

\begin{figure}
\centering
\includegraphics[width=0.47\textwidth,page=1]{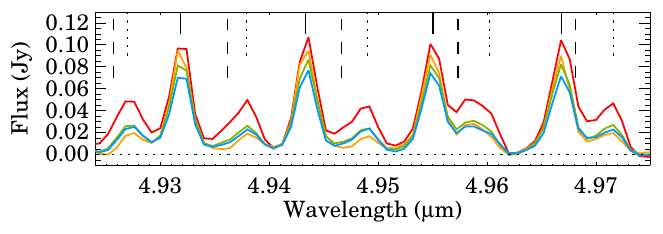}
\caption{Fundamental CO lines in the spectra of DQ\,Tau. Vertical solid / dotted / dashed lines mark $v=1-0$ / $2-1$ / $3-2$ transitions, respectively.}\label{fig:co_comparison}
\end{figure}

The fundamental rovibrational CO lines also show strong variability with time (Fig.~\ref{fig:co_comparison}). To learn about how physical conditions change in the emitting CO gas in the disk of DQ\,Tau, we used simple slab models to calculate a grid with temperatures ($T$) between 100\,K and 4000\,K with increments of 50\,K and column densities ($N$) between $10^{17}$\,m$^{-2}$ and $10^{24}$\,m$^{-2}$ with increments of ${\log_{10}}N=0.1$, using the slabspec software \citep{salyk2020}. We then scaled the slab models by changing the emitting area until the $\chi^2$ calculated for the $4.9-5.1\,\mu$m wavelength range became minimal. This was done for each $T$ and $N$ pair. The emitting area is assumed to be circular with a radius of $R$. Following \citet{colmenares2024}, we calculated the $\chi^2_{\rm min}/\chi^2$ surface for the whole grid and found the best-fitting model by calculating the centroid of this surface. These are plotted in Fig.~\ref{fig:co} in the Appendix, along with the 1$\sigma$, 2$\sigma$, 3$\sigma$ contours. In case of CO, we found that the emission is optically thin and therefore $N$ and the emitting area are degenerate, as indicated by the open and elongated 1$\sigma$ contours. Therefore, while the temperature is well-determined, we could only give an upper limit for the column density of CO. As a consequence, the calculated radii of the emitting area are lower limits. At $2261\pm46$\,K, the temperature of CO is significantly higher in epoch 1 than in epoch 2 ($1553\pm18$\,K), while in epochs 3-4, the CO temperature was in-between ($1875\pm36$\,K and $1867\pm28$\,K, respectively). Upper limits for the column density are in the $N=10^{20.24} - 10^{20.70}$\,m$^{-2}$ range. The exact values are listed in Tab.~\ref{tab:slab}. The best-fitting CO slab models are plotted over the observed spectra in Fig.~\ref{fig:co}.

To test how robust these results are and to search for a better fit, we tried to model the CO lines with a disk model with a power law temperature gradient. We fixed the power law exponent to $-0.75$; keeping it free did not improve the fits. We fixed the column density to $N=10^{21}$\,m$^{-2}$. We then constructed a disk model from adjacent rings of decreasing temperature with $T_{\rm in}$ being the temperature at the inner radius and $T_{\rm out}$ being the temperature at the outer radius. We found that the best-fitting temperatures are
$T_{\rm in} = 4000$\,K and $T_{\rm out} = 1400$\,K in epoch 1,
$T_{\rm in} = 2700$\,K and $T_{\rm out} = 750$\,K in epoch 2,
$T_{\rm in} = 4000$\,K and $T_{\rm out} = 900$\,K in epoch 3, and
$T_{\rm in} = 4000$\,K and $T_{\rm out} = 900$\,K in epoch 4.
The maximal temperatures in most cases are not well-determined because $4000$\,K is the higher end of our grid but the emitting area of the rings with the highest temperatures in our model are very small, therefore they do not contribute much flux to the total model. The lowest temperatures, however, indicate similar time-evolution to the single temperature slab models: hottest in the first epoch, coldest in the second epoch, and intermediate in the last two epochs.

\begin{figure}
\centering
\includegraphics[width=0.48\textwidth,page=2]{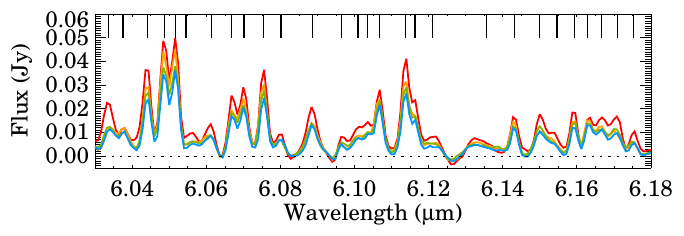}\\
\includegraphics[width=0.45\textwidth,page=4]{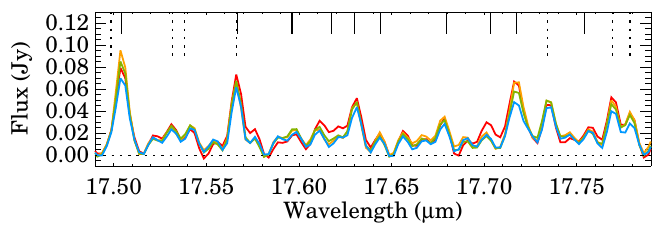}\\
\includegraphics[width=0.45\textwidth,page=5]{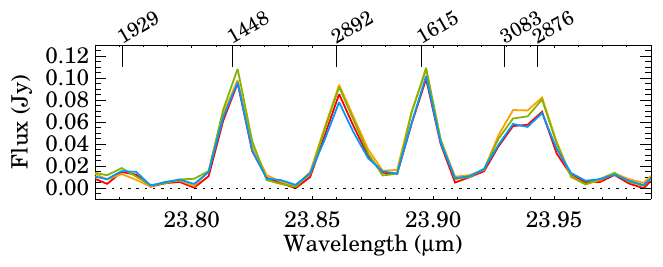}
\caption{Upper panel: rovibrational H$_2$O lines in the spectra of DQ\,Tau. Middle panel: rotational H$_2$O (solid marks) and OH (dotted marks) lines in the spectra of DQ\,Tau. Lower panel: rotational H$_2$O lines (solid marks) with their upper level energies in K indicated above the panel.}\label{fig:h2o_oh_comparison}
\end{figure}

We also used single temperature slab models to fit the H$_2$O lines in the spectra of DQ\,Tau. We considered the $5.3-6.3\,\mu$m range, which is dominated by rovibrational H$_2$O lines, the $13.4-15.1\,\mu$m range, where a water model is necessary to be able to fit the HCN and CO$_2$ features, and the $17-25\,\mu$m range, dominated by pure rotational lines. Here, we calculated grids for temperatures between 200 and 1450\,K with increments of 25\,K and column densities between $10^{17}$ and $10^{26}$\,m$^{-2}$ with increments of ${\log_{10}}N=0.25$. As opposed to CO, for H$_2$O both the temperature and the column density are well-determined, because most water lines are optically thick. The obtained values are displayed in Tab.~\ref{tab:slab}. The rovibrational lines at $5.3-6.3\,\mu$m  trace hotter water in the $738 - 855$\,K range. This water component follows similar time evolution to CO: the obtained temperature is higher in epoch 1, lower in epoch 2, and intermediate temperatures are obtained for epochs 3-4, although the differences are not as significant as for CO. The rotational lines in the $17-25\,\mu$m range are somewhat colder, in the $621 - 664$\,K range, and they show no significant time-variability: the temperature in the first three epochs agree within 1$\sigma$, although there is a 2$\sigma$ hint for a lower temperature in epoch 4. Column densities do not change much with time, and the rotational lines seem to trace higher column densities than the rovibrational lines. Selected H$_2$O lines from these wavelength ranges are displayed in Fig.~\ref{fig:h2o_oh_comparison}. The bottom panel of this figure shows two specific H$_2$O lines with low upper level energies: one at $23.817\,\mu$m with $E_{\rm up} = 1448\,$K and one at $23.895\,\mu$m with $E_{\rm up} = 1615\,$K. These lines trace cold water, just after sublimation from the ice phase near the snowline. These lines are clearly detected in DQ\,Tau and the 1448/1615 line ratio suggests the presence of an even colder, $170 - 200\,$K H$_2$O component around the water snowline \citep[cf.][]{banzatti2025,smith2025}. These lines appear mostly constant during our monitoring, suggesting that this cold water component did not react much to the accretion luminosity changing from 0.1 to 0.4\,L$_{\odot}$ on a few weeks timescale.

To fit the CO$_2$ and HCN features, we first subtracted the best-fitting H$_2$O slab model in the $13.4-15.1\,\mu$m wavelength range and fitted the CO$_2$ and HCN models to the residuals. The temperature and column density grid for these molecules were the same as for H$_2$O. We calculated $\chi^2$ in the $4.91 - 15.0\,\mu$m range for CO$_2$ and in the $13.8 - 14.1\,\mu$m range for HCN. The values obtained for the best-fitting models are listed in Tab.~\ref{tab:slab}. CO$_2$ seems to trace the coldest gas from the detected molecules in the $520-624$\,K range. Interestingly, it has similar temperatures in epochs 1-2 and coldest in epoch 3. HCN traces hotter gas in the $878 - 967$\,K range. It shows the highest temperatures in epochs 1 and 4, lowest in epoch 2, and an intermediate value in epoch 3, correlating most closely with the changes in accretion luminosity. Both CO$_2$ and HCN are optically thin, but only upper limits for the HCN column densities (and lower limits for the radii of the emitting area) could be given. Fig.~\ref{fig:co2hcn} shows the H$_2$O, CO$_2$, and HCN models as well as their sum plotted over the observed spectra. We searched for C$_2$H$_2$ emission in the residuals, but it remains undetected.

We detect rotational OH lines throughout the MIRI spectra of DQ\,Tau from 10 to 26$\,\mu$m. These lines have a wide range of upper level energies from 26\,800\,K at 10.2$\,\mu$m to 3500\,K at 25.2$\,\mu$m, indicating the presence of OH gas with a significant temperature gradient. Transitions with the highest upper level energies are likely the result of the photodissociation of H$_2$O by UV photons. We do not attempt to model these lines at this time, but a visual comparison of the spectra hints for variability. Lines with $E_{\rm up} \gtrsim 7000\,$K are strongest in epoch 1 (where accretion was strongest), while lines with $E_{\rm up} \lesssim 7000\,$K are strongest in epoch 2 (apastron). This suggest that the increased accretion may lead to stronger photodissociation.

\begin{table*}
\caption{Wavelength ranges for fitting the various molecular features and the obtained excitation temperatures ($T$), column densities ($N$) and slab radii ($R$) from our slab modeling in Sec.\ref{sec:gaslines}.}
\label{tab:slab}
\centering
\footnotesize
\begin{tabular}{l c c c c c c c}
\hline\hline
Molecule                    & CO          & H$_2$O         & H$_2$O & H$_2$O & CO$_2$ & HCN  \\
\hline
Fitting range ($\mu$m)      & $4.9 - 5.1$ & $5.3 - 6.3$    & $13.4 - 15.1$  & $17.0 - 25.0$  & $14.9 - 15.0$  & $13.8-14.1$ \\
Epoch 1 $T$ (K)             & $2261\pm46$ & $855\pm31$     & $679\pm5$      & $664\pm16$     & $624\pm4$      & $967\pm25$  \\
Epoch 2 $T$ (K)             & $1553\pm18$ & $738\pm19$     & $678\pm3$      & $653\pm10$     & $615\pm11$     & $878\pm14$  \\
Epoch 3 $T$ (K)             & $1875\pm36$ & $785\pm22$     & $694\pm3$      & $648\pm12$     & $520\pm9$      & $922\pm24$  \\
Epoch 4 $T$ (K)             & $1867\pm28$ & $773\pm26$     & $631\pm2$      & $621\pm15$     & $560\pm3$      & $977\pm29$  \\
Epoch 1 log$_{10}N$ (m$^{-2}$) & $<20.28$ & $22.22\pm0.01$ & $22.53\pm0.01$ & $22.87\pm0.04$ & $20.65\pm0.01$ & $<19.67$    \\
Epoch 2 log$_{10}N$ (m$^{-2}$) & $<20.70$ & $22.01\pm0.01$ & $22.29\pm0.01$ & $22.56\pm0.02$ & $20.38\pm0.01$ & $<19.41$    \\
Epoch 3 log$_{10}N$ (m$^{-2}$) & $<20.38$ & $22.06\pm0.01$ & $22.48\pm0.01$ & $22.60\pm0.03$ & $20.66\pm0.01$ & $<19.90$    \\
Epoch 4 log$_{10}N$ (m$^{-2}$) & $<20.24$ & $22.07\pm0.01$ & $22.52\pm0.01$ & $22.79\pm0.03$ & $20.69\pm0.01$ & $<19.73$    \\
Epoch 1 R (au)              & $>0.60$     & $0.26\pm0.02$  & $0.69\pm0.01$  & $0.65\pm0.03$  & $0.42\pm0.01$  & $>0.94$     \\
Epoch 2 R (au)              & $>0.47$     & $0.35\pm0.02$  & $0.80\pm0.01$  & $0.77\pm0.02$  & $0.54\pm0.01$  & $>1.22$     \\
Epoch 3 R (au)              & $>0.53$     & $0.28\pm0.01$  & $0.65\pm0.01$  & $0.74\pm0.02$  & $0.52\pm0.01$  & $>0.62$     \\
Epoch 4 R (au)              & $>0.59$     & $0.28\pm0.02$  & $0.71\pm0.01$  & $0.69\pm0.03$  & $0.46\pm0.01$  & $>0.70$     \\
\hline
\end{tabular}
\end{table*}

\section{Discussion}
\label{sec:discussion}

\subsection{DQ\,Tau in the context of JWST's view of T Tauri disks}

As we described in the previous section, most of the line features present in DQ\,Tau's MIRI spectra in Fig.~\ref{fig:comparison} are from water, while C$_2$H$_2$ and other hydrocarbons are absent. This suggests that the disk of DQ\,Tau is water-rich but unusually hydrocarbon-poor. To put these results into context more quantitatively, we compare DQ\,Tau to the JDISC sample, 31 T Tauri disks around $0.1-2\,$M$_{\odot}$ stars observed with JWST/MIRI \citep{pontoppidan2024}. \citet{arulanantham2025} studied the gas line emission in this sample by fitting slab models and calculating line luminosities for HCN, CO$_2$, C$_2$H$_2$, and H$_2$O. For the former three molecules, because they are heavily blended with each other and with water, they integrated the total luminosity of their fitted slab models between $12-16\,\mu$m. For H$_2$O, they used two cold water tracer lines at $23.82\,\mu$m  and $23.90\,\mu$m. They calculated the luminosity ratios $L_{\rm C2H2} / L_{\rm CO2}$ and $L_{\rm HCN} / L_{\rm H2O,cold}$ and plotted these values for their sample as a function of the millimeter dust disk size. For the latter, they used boundaries that contain 90\% to 95\% of the flux at 1.3\,mm.

\begin{figure}
\centering
\includegraphics[width=\columnwidth]{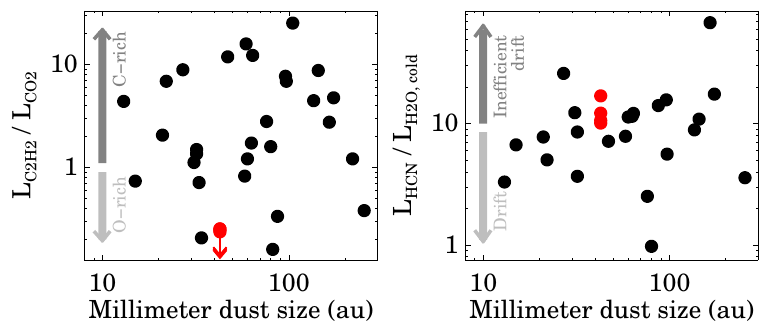}
\caption{Line luminosity ratios as a function of millimeter dust size for the JDISC sample (black dots, from \citealt{arulanantham2025}) and for DQ\,Tau (red symbols).}\label{fig:nicole}
\end{figure}

In Fig.~\ref{fig:nicole}, we reproduced Fig.~17 of \citet{arulanantham2025} and added DQ\,Tau by deriving the necessary line luminosity ratios in the same way. For the millimeter dust size, we used a 95\% size of 42\,au determined by \citet{long2019} from 1.3\,mm ALMA observations. Fig.~\ref{fig:nicole} (left) shows that DQ\,Tau (marked with red downward arrow to indicate upper limits) is indeed among the most O-rich / C-poor disks. Fig.~\ref{fig:nicole} (right) shows a weak trend between the millimeter dust size and the $L_{\rm HCN} / L_{\rm H2O,cold}$ ratio, explained by large disks having significant substructures (gaps and rings) that halt the inward drift of icy pebbles leading to weaker cold H$_2$O emission \citep{banzatti2025}. Here, DQ\,Tau fits into the trend and its luminosity ratio is consistent with having a marginally inefficient pebble drift in its disk. We used all four MIRI spectra to calculate the $L_{\rm HCN} / L_{\rm H2O,cold}$ ratio and found four distinct values. This is because while the cold water emission was constant, HCN fluxes varied due to the accretion rate changes (see below). This suggests that the large scatter and weak trend in Fig.~\ref{fig:nicole} (right) may be the consequence of accretion variability.

\begin{figure}
\centering
\includegraphics[width=\columnwidth]{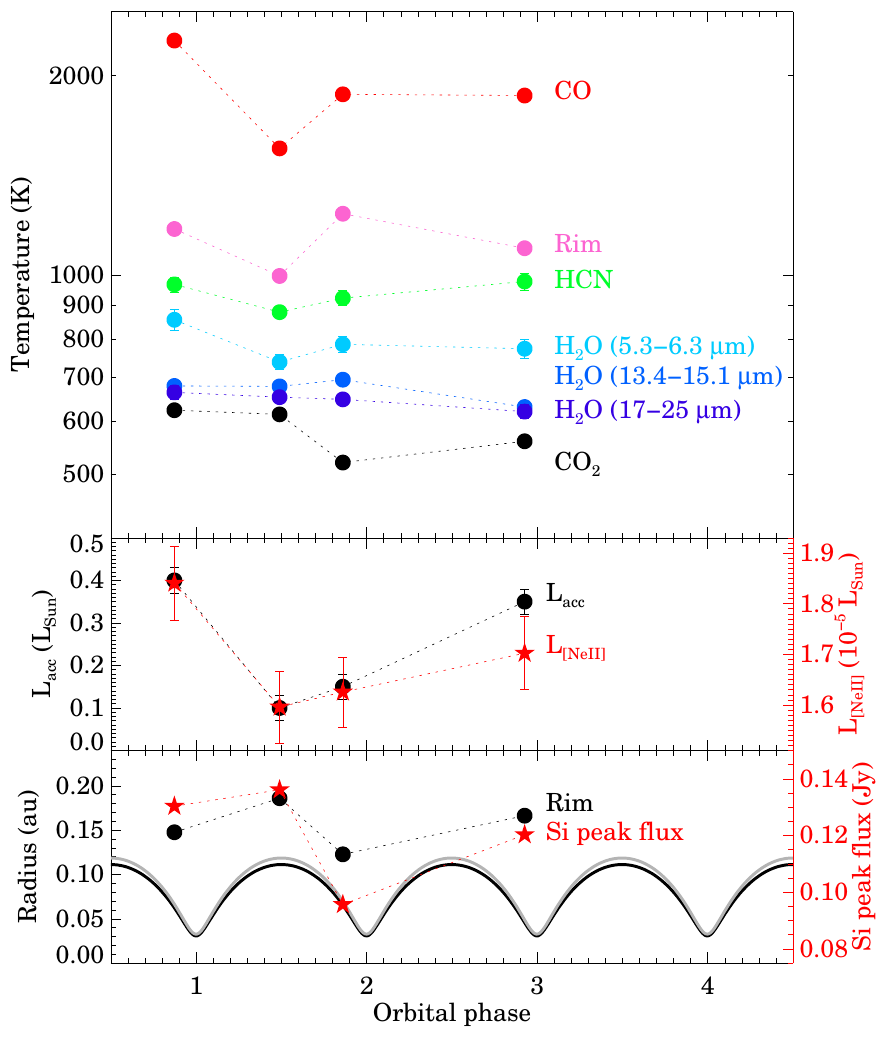}
\caption{Temperature of the dust rim and of various gas components (top), accretion luminosity and the luminosity of the [Ne\,II] line at 12.81$\,\mu$m (middle), dust rim radius, peak flux of the continuum-subtracted 10$\,\mu$m silicate feature, and the distance of the primary (black) and secondary (gray) from the barycenter (bottom). Error bars smaller than the symbol size are not plotted.}\label{fig:temperature}
\end{figure}

\begin{figure}
\centering
\includegraphics[width=\columnwidth]{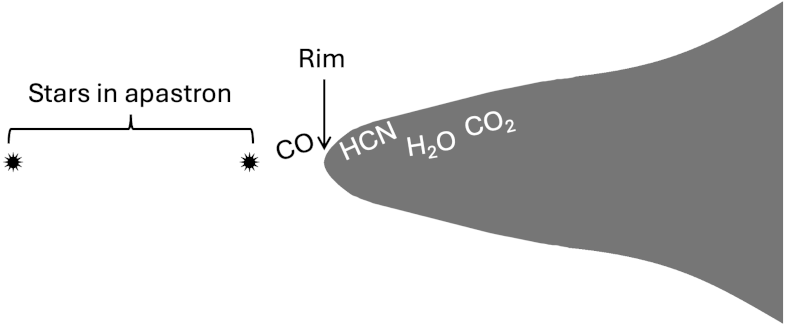}\\
\includegraphics[width=\columnwidth]{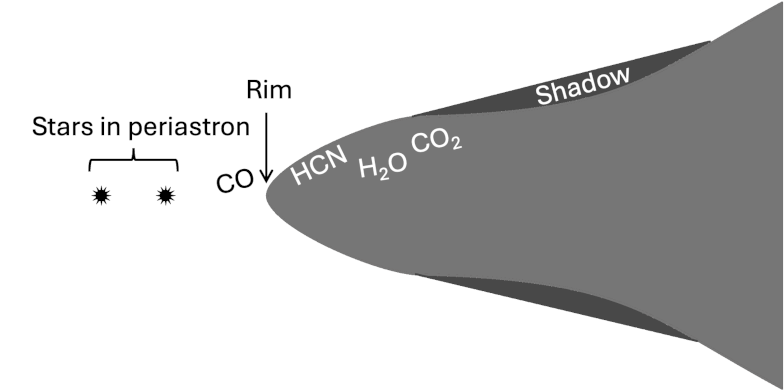}
\caption{Simplistic sketch of the DQ\,Tau system during apastron (top) and periastron (bottom).}\label{fig:sketch}
\end{figure}

Fig.~\ref{fig:temperature} summarizes the temperatures of various gas components obtained from our slab modeling (Sec.~\ref{sec:gaslines}) and if the dust rim (Sec.~\ref{sec:sed}). CO traces the hottest gas with temperatures  between $1553-2261\,$K, then comes the HCN between $878-977\,$K, then the hot  H$_2$O between $738-855\,$K, then the warm H$_2$O between $615-664\,$K, and finally CO$_2$ is the coldest between $520-624\,$K. This is in accordance with the findings of \citet{arulanantham2025}, who found that C$_2$H$_2$ and HCN emission typically traces hotter regions (median temperatures in the JDISC sample are 920\,K and 820\,K, respectively) than CO$_2$ (median temperature of 600\,K). In this respect, the disk of DQ\,Tau does not differ from a typical T\,Tauri disk. While we found mostly hot and warm water in DQ\,Tau, our warm water model does not fit very well the H$_2$O lines at $23.82\,\mu$m and $23.90\,\mu$m, which have low upper level energies ($1448-1615\,$K, \citealt{banzatti2025}) and agree more with water at $\sim400\,$K.

CO is hotter than the dust sublimation temperature ($\sim1500\,$K for astrosilicates, e.g., \citealt{pollack1994,isella2005,kobayashi2011}) in all epochs, therefore the bulk of the CO emission originates either from the inner dust-free gaseous disk or from the hot material in the accretion columns. Emission from other molecules comes from the dusty part of the disk, beyond the dust rim. This is illustrated in a simplistic sketch of the DQ\,Tau system in Fig.~\ref{fig:sketch}. The temperatures we obtain for the dust rim are in the range of $996-1237\,$K, lower than the dust sublimation temperature. This suggests that the inner rim is farther out than it would be under the influence of radiation only, pointing to a larger dust-free inner hole, probably shaped by the dynamical effects of the central eccentric binary.

\subsection{Time-variable gas and dust emission in DQ\,Tau}

Fig.~\ref{fig:temperature} visualizes the time-evolution of the temperatures of the various gas and dust components along with the accretion luminosity, allowing us to search for correlations between these quantities. There is no one-on-one correlation between the plotted temperatures and the accretion luminosity. Nevertheless, the temperature of CO, the dust rim, HCN, and the hot water component seem to be reacting most strongly to accretion luminosity changes. These molecules trace hotter gas, presumably closer to the accreting stars, therefore, their temperatures are more strongly affected by the accretion heating. The warm water component is more constant in time. This emission traces more distant disk regions ($0.7-0.8\,$au for the warm water versus $0.3\,$au for the hot water), where changes in the radiation field from accretion seem to be less important. The cold water component close to the snowline, as traced by the luminosity of the low upper level energy H$_2$O lines at $23.817\,\mu$m and $23.895\,\mu$m also appear to be mostly constant. The middle panel of Fig.~\ref{fig:temperature} shows that there is one quantity that correlates very tightly with the accretion luminosity changes, the luminosity of the forbidden [Ne\,II] line at $12.81\,\mu$m. As we mentioned in Sec.~\ref{sec:gaslines}, this line traces the ionized disk surface in DQ\,Tau. During epochs of higher accretion rate, the disk probably receives a higher flux of ionizing radiation, explaining this effect.

To better understand the various processes in DQ Tau's disk, we can investigate the timescales for photodissociation, thermal sublimation, and freeze-out in the inner disk in relation to the orbital period of the stars. Photodissociation in the UV-exposed surface layer (that we can observe using molecular line emission) is very quick: tens to hundreds of seconds in case of unshielded conditions \citep{visser2009, walsh2012}. This is faster than mixing and formation timescales of the various molecules. Mixing depends on the radius and the turbulent $\alpha$ parameter, but for typical alpha values of $10^{-3} - 10^{-2}$, its timescale is measured in years or decades even at 1\,au or closer. Therefore, steady state requires a continuous replenishment of molecules from deeper layers. In DQ\,Tau with its rapidly changing radiation field and temperature, molecules are probably replenished on the disk surface by fast sublimation or photodesorption processes. The thermal relaxation timescale of the dust grains are also fast. At the water snowline ($T\sim170\,$K) and grain sizes typical for protoplanetary disks, dust grains in the surface layer adjust their temperature to the changing radiation field on the order of seconds to hours \citep{cg97}. Once the temperature is above the water sublimation temperature, small grains quickly lose their ice mantles, again on the order of minutes to hours (see desorption timescales in \citealt{fraser2001}). Freeze-out is much slower and can take decades \citep{houge2023, rab2017a}. This suggests that while hotter gas such as CO, the hot H$_2$O component, and HCN react quickly to changes in accretion rate, the cold H$_2$O component close to the snowline would be much slower to react. In a system with a periodically changing accretion rate like DQ Tau, where the timescale of accretion rate changes is about two orders of magnitude shorter than the freeze-out timescale, no significant changes are expected in the cold water abundance on days-to-weeks timescale. These are in line with our observations.

To study the variability of the dust rim, we took the total (stellar + accretion) luminosity of the system in the four MIRI epochs ($1.72\,$L$_{\odot}$, $1.42\,$L$_{\odot}$, $1.47\,$L$_{\odot}$, and $1.67\,$L$_{\odot}$, respectively), and assuming that the radiation field weakens by the square of the distance from the stars, we calculated the distance from the barycenter where dust has the temperature we calculated for the rim ($1173\,$K, $996$\,K, $1237$\,K, and $1097$\,K). The obtained distances are plotted in the bottom panel of Fig.~\ref{fig:temperature}. This panel also shows the distances of the two stellar components from the barycenter. We find that the rim is closer to the barycenter at the periastron epochs and farther from it at apastron. This suggests a very dynamic inner disk, which pulsates with the binary orbit. \citet{sytov2011} found in their two-dimensional gas-dynamical numerical simulations that the inner radius of the disk around DQ\,Tau fluctuates about its mean value with a period of about $5P_{\rm orb}$, but their rim is also farther from the barycenter, at 0.45\,au. Our results indicate that the dust rim is closer and its location may vary with a period of $P_{\rm orb}$ instead. We note that we have no observational constraint on the spatial structure of the dust rim, therefore it may not be a stable structure in the disk, but may rather be some dusty streams of material that are present only temporarily and accreted onto the stars later.

The rim radii we find are smaller than the rim radii calculated around an eccentric binary from hydrodynamical simulations \citep[$0.35-0.45\,$au][]{gunther2002,sytov2011}, suggesting that the dust we found emitting at $996-1173$\,K temperatures are indeed not stable structures. Our results show that both dusty and gaseous (CO) material exist within the inner gap. This is not surprising, because the numerical simulations show that material is flowing through the gap and is accreted onto the stars, and observations are also indicating non-negligible accretion rates in DQ\,Tau. There were earlier indications for the presence of such material: \citet{boden2009} found that they needed to include a spatially resolved near-infrared excess source to model the $K$-band visibilities obtained with the Keck Interferometer. They found that this source is extended on a 0.2\,au scale at apastron, very similar to the rim radius we measure for epoch 2 (apastron).

\subsection{Time-variable shadowing in the DQ\,Tau disk}

DQ\,Tau displays an interesting relation between the strength of the silicate feature and the rim properties that suggest time-variable shadowing: at apastron, in epoch 2, the rim is coldest and farthest from the barycenter, therefore, the inner rim may not be very puffed-up and shadowing is minimal, leading to a strong $10\,\mu$m silicate feature. Close to periastron, in epoch 3, the rim is hottest and closest to the barycenter, so the inner rim may be very puffed-up and may cast a significant shadow onto the outer disk. Thus, the $10\,\mu$m micron silicate feature is the weakest in this epoch. This is supported by the correlation of the rim radius and the peak flux of the continuum-subtracted silicate feature plotted in the bottom panel of Fig.~\ref{fig:temperature}. This idea is illustrated in Fig.~\ref{fig:sketch}.

In Sec.~\ref{sec:sed} we calculated the luminosity of the rim in the four studied epochs (0.64\,L$_{\odot}$, 0.31\,L$_{\odot}$, 0.52\,L$_{\odot}$, and 0.40\,L$_{\odot}$, respectively). We theorize that the rim is heated by the central binary components and the accretion onto them, therefore, the luminosity of the rim should depend on the stellar+accretion luminosity in that moment and the distance of the rim from the stellar components (as both the stellar and the accretion radiation originates at the surfaces of the stars). Indeed, there is again a correlation between the rim luminosity and the stellar+accretion luminosity divided by the square of the star-rim distance. This means that the rim captures the time-variable energy released by the central accreting sources and within a certain area its dust and gas content heats up as a result, while farther away disk regions are either unchanged or affected by the variable shadowing.

\section{Conclusions and summary}

With the goal to investigate how moderate, 1--2 mag brightenings of T\,Tauri stars affect the structure and composition of their protoplanetary disks, we executed an experiment utilizing the predictable periodic accretion pulses of DQ\,Tau driven by its binary nature. We conducted a four-epoch monitoring with JWST/MIRI, supplemented by optical and infrared photometric and spectroscopic measurements. Our data confirm that DQ Tau brightened close to the predicted periastron epochs, reaching accretion luminosities up to 1\,L$_{\odot}$.

We modeled multi-epoch SEDs with stellar, accretion and dusty rim components. From the rim temperature, we derived a time-dependent rim radius of $0.12 - 0.19\,$au, smaller than the $0.35-0.45\,$ au cavity size predicted by dynamical simulations. We found that the rim radius pulsates with the binary orbital period. Exploring the gas content of the inner disk, we detected atomic H\,I, ionized [Ar\,II] and [Ne\,II], hot CO and OH, warm HCN, a multi-temperature H$_2$O reservoir, and cold CO$_2$. Comparing with a larger sample of T\,Tauri disks from the JWST JDISC survey, molecular line ratios imply that DQ\,Tau has an O-rich / C-poor disk, with somewhat inefficient pebble drift. The high temperature derived for CO suggests the presence of gaseous material within the inner gap, consistently with the general picture of accretion from a circumbinary disk. Since the inner rim is colder than the dust sublimation temperature, we may assume that dusty material may also move inward temporarily. Thus the inner edge of the disk may be formed by a time-dependent distribution of dust rather than a well-defined, stable rim structure.  

The disk area behind the rim contains atomic and molecular gas but also emits solid-state features from silicates. Fitting its profile revealed a mineralogy dominated by large amorphous olivines, pyroxenes, and crystalline enstatites which were constant in time. The strength of the silicate feature, however, is variable, being smallest when the rim is closest to the stars, and highest when the rim is further from the stars, suggesting that we witness variable shadowing in action. Our results on DQ\,Tau demonstrate that moderate amplitude accretion variability can modify the structure and physical properties of the inner disk, affecting the birthplace of terrestrial planets. For future studies of accretion-related phenomena and the effects of accretion variability on the disk, we emphasize the importance of well-timed, coordinated, multi-wavelength efforts that cover with spectroscopic instruments as much of the electromagnetic spectrum as possible.

\begin{acknowledgements}
We are grateful to the referee for helpful comments that improved our manuscript. We thank the observers at La Silla, Maren Hempel, Angela Hempel, and Sim\'on \'Angel for taking the FEROS spectra for us. We thank those who observed DQ\,Tau and uploaded their data to the AAVSO International Database: 
Kevin Alton,
Enrique T. Boeneker,
Stephen M. Brincat,
Andrew Corkill, 
William Dillon,
Charles Galdies,
Sauro Gaudenzi,
Franz-Josef (Josch) Hambsch,
Thomas Haugh,
David Iadevaia,
Thomas Karlsson,
Matthias Kolb,
Peter Lindner,
James (Bruce) McMath, 
Gordon Myers,
Kenneth Menzies,
Matthew Patrick,
Ivo Peretto,
Michael Poxon, 
Thomas Rutherford,
Ian Sharp,
Andre Steenkamp,
Robert Stephens,
Ray Tomlin,
Bill Tschumy, and
Mike Wiles. Their enthusiasm and tireless work at a short notice was a great inspiration for this project. This work is based in part on observations made with the NASA/ESA/CSA James Webb Space Telescope. The data were obtained from the Mikulski Archive for Space Telescopes at the Space Telescope Science Institute, which is operated by the Association of Universities for Research in Astronomy, Inc., under NASA contract NAS 5-03127 for JWST. These observations are associated with program \#4876. The specific observations analyzed can be accessed via \href{http://dx.doi.org/10.17909/pkx6-3g49}{doi:10.17909/pkx6-3g49}. Support for program \#4876 was provided by NASA through a grant from the Space Telescope Science Institute, which is operated by the Association of Universities for Research in Astronomy, Inc., under NASA contract NAS 5-03127. This research is based in part on observations made with the Neil Gehrels Swift Observatory, obtained from the MAST data archive at the Space Telescope Science Institute, which is operated by the Association of Universities for Research in Astronomy, Inc., under NASA contract NAS 5–26555. This work is based in part on observations collected at the European Southern Observatory under ESO programme 0114.C-6020(A) or 114.28H7.001. The IRTF/SpeX data presented here were obtained at the Infrared Telescope Facility, which is operated by the University of Hawaii under contract 80HQTR24DA010 with the National Aeronautics and Space Administration. Our telescope time on REM was funded by the OPTICON-RadioNet Pilot (ORP) optical transnational access program. ORP is supported by the European Union’s Horizon 2020 research and innovation programme under grant agreement No 101004719. The operation of the RC80 telescope at Konkoly Observatory has been supported by the GINOP 2.3.2-15-2016-00033 grant of the National Research, Development and Innovation Office (NKFIH) funded by the European Union. This work was also supported by the NKFIH NKKP grant ADVANCED 149943, the NKFIH excellence grant TKP2021-NKTA-64, and the NKFIH \'Elvonal grants KKP-143986 and KKP-137523 `SeismoLab'. Project no.149943 has been implemented with the support provided by the Ministry of Culture and Innovation of Hungary from the National Research, Development and Innovation Fund, financed under the NKKP ADVANCED funding scheme. This work received funding from the Hungarian NKFIH OTKA project no.~K-147380. D.~S. was funded by the Deutsche Forschungsgemeinschaft (DFG, German Research Foundation) – project number: 550639632. B.~P-R.~acknowledges support by NASA STScI grant No.~JWST-GO-04876. This research was supported in part by grant NSF PHY-2309135 to the Kavli Institute for Theoretical Physics (KITP). 
\end{acknowledgements}

%\bibliographystyle{aa}
%\bibliography{main}

\begin{thebibliography}{99}
\expandafter\ifx\csname natexlab\endcsname\relax\def\natexlab#1{#1}\fi

\bibitem[{{{\'A}brah{\'a}m} {et~al.}(2019){{\'A}brah{\'a}m}, {Chen},
  {K{\'o}sp{\'a}l}, {Bouwman}, {Carmona}, {Haas}, {Sicilia-Aguilar}, {Sobrino
  Figaredo}, {van Boekel}, \& {Varga}}]{abraham2019}
{{\'A}brah{\'a}m}, P., {Chen}, L., {K{\'o}sp{\'a}l}, {\'A}., {et~al.} 2019,
  \apj, 887, 156

\bibitem[{{{\'A}brah{\'a}m} {et~al.}(2009){{\'A}brah{\'a}m}, {Juh{\'a}sz},
  {Dullemond}, {K{\'o}sp{\'a}l}, {van Boekel}, {Bouwman}, {Henning},
  {Mo{\'o}r}, {Mosoni}, {Sicilia-Aguilar}, \& {Sipos}}]{abraham2009}
{{\'A}brah{\'a}m}, P., {Juh{\'a}sz}, A., {Dullemond}, C.~P., {et~al.} 2009,
  \nat, 459, 224

\bibitem[{{Alcal{\'a}} {et~al.}(2017){Alcal{\'a}}, {Manara}, {Natta}, {Frasca},
  {Testi}, {Nisini}, {Stelzer}, {Williams}, {Antoniucci}, {Biazzo}, {Covino},
  {Esposito}, {Getman}, \& {Rigliaco}}]{alcala2017}
{Alcal{\'a}}, J.~M., {Manara}, C.~F., {Natta}, A., {et~al.} 2017, \aap, 600,
  A20

\bibitem[{{Alexander} {et~al.}(2014){Alexander}, {Pascucci}, {Andrews},
  {Armitage}, \& {Cieza}}]{alexander2014}
{Alexander}, R., {Pascucci}, I., {Andrews}, S., {Armitage}, P., \& {Cieza}, L.
  2014, in Protostars and Planets VI, ed. H.~{Beuther}, R.~S. {Klessen}, C.~P.
  {Dullemond}, \& T.~{Henning}, 475--496

\bibitem[{{Artymowicz} \& {Lubow}(1994)}]{artymowicz1994}
{Artymowicz}, P. \& {Lubow}, S.~H. 1994, \apj, 421, 651

\bibitem[{{Artymowicz} \& {Lubow}(1996)}]{artymowicz1996}
{Artymowicz}, P. \& {Lubow}, S.~H. 1996, \apjl, 467, L77

\bibitem[{{Arulanantham} {et~al.}(2025){Arulanantham}, {Salyk}, {Pontoppidan},
  {Banzatti}, {Zhang}, {{\"O}berg}, {Long}, {Carr}, {Najita}, {Pascucci},
  {Jos{\'e} Colmenares}, {Xie}, {Huang}, {Green}, {Andrews}, {Blake}, {Bergin},
  {Pinilla}, {Vioque}, {Dahl}, {Raul}, {Krijt}, \& {the JDISCS
  Collaboration}}]{arulanantham2025}
{Arulanantham}, N., {Salyk}, C., {Pontoppidan}, K., {et~al.} 2025, arXiv
  e-prints, arXiv:2505.07562

\bibitem[{{Ballering} \& {Eisner}(2019)}]{ballering2019}
{Ballering}, N.~P. \& {Eisner}, J.~A. 2019, \aj, 157, 144

\bibitem[{{Banzatti} {et~al.}(2012){Banzatti}, {Meyer}, {Bruderer}, {Geers},
  {Pascucci}, {Lahuis}, {Juh{\'a}sz}, {Henning}, \&
  {{\'A}brah{\'a}m}}]{banzatti2012}
{Banzatti}, A., {Meyer}, M.~R., {Bruderer}, S., {et~al.} 2012, \apj, 745, 90

\bibitem[{{Banzatti} {et~al.}(2015){Banzatti}, {Pontoppidan}, {Bruderer},
  {Muzerolle}, \& {Meyer}}]{banzatti2015}
{Banzatti}, A., {Pontoppidan}, K.~M., {Bruderer}, S., {Muzerolle}, J., \&
  {Meyer}, M.~R. 2015, \apjl, 798, L16

\bibitem[{{Banzatti} {et~al.}(2025){Banzatti}, {Salyk}, {Pontoppidan}, {Carr},
  {Zhang}, {Arulanantham}, {Krijt}, {{\"O}berg}, {Cleeves}, {Najita},
  {Pascucci}, {Blake}, {Romero-Mirza}, {Bergin}, {Cieza}, {Pinilla}, {Long},
  {Mallaney}, {Xie}, {Waggoner}, {Kaeufer}, \& {The Jdiscs
  Collaboration}}]{banzatti2025}
{Banzatti}, A., {Salyk}, C., {Pontoppidan}, K.~M., {et~al.} 2025, \aj, 169, 165

\bibitem[{{Boden} {et~al.}(2009){Boden}, {Akeson}, {Sargent}, {Carpenter},
  {Ciardi}, {Bary}, \& {Skrutskie}}]{boden2009}
{Boden}, A.~F., {Akeson}, R.~L., {Sargent}, A.~I., {et~al.} 2009, \apjl, 696,
  L111

\bibitem[{{Bosman} {et~al.}(2017){Bosman}, {Bruderer}, \& {van
  Dishoeck}}]{bosman2017}
{Bosman}, A.~D., {Bruderer}, S., \& {van Dishoeck}, E.~F. 2017, \aap, 601, A36

\bibitem[{{Bruderer} {et~al.}(2015){Bruderer}, {Harsono}, \& {van
  Dishoeck}}]{bruderer2015}
{Bruderer}, S., {Harsono}, D., \& {van Dishoeck}, E.~F. 2015, \aap, 575, A94

\bibitem[{{Calahan} {et~al.}(2024){Calahan}, {Bergin}, {van't Hoff}, {Booth},
  {{\"O}berg}, {Zhang}, {Calvet}, \& {Hartmann}}]{calahan2024}
{Calahan}, J.~K., {Bergin}, E.~A., {van't Hoff}, M., {et~al.} 2024, \apj, 975,
  170

\bibitem[{{Cardelli} {et~al.}(1989){Cardelli}, {Clayton}, \&
  {Mathis}}]{cardelli1989}
{Cardelli}, J.~A., {Clayton}, G.~C., \& {Mathis}, J.~S. 1989, \apj, 345, 245

\bibitem[{{Cecil} \& {Flock}(2024)}]{cecil2024}
{Cecil}, M. \& {Flock}, M. 2024, \aap, 692, A171

\bibitem[{{Chiang} \& {Goldreich}(1997)}]{cg97}
{Chiang}, E.~I. \& {Goldreich}, P. 1997, \apj, 490, 368

\bibitem[{{Christiaens} {et~al.}(2023){Christiaens}, {Gonzalez}, {Farkas},
  {Dahlqvist}, {Nasedkin}, {Milli}, {Absil}, {Ngo}, {Cantero}, {Rainot},
  {Hammond}, {Bonse}, {Cantalloube}, {Vigan}, {Kompella}, \&
  {Hancock}}]{Christiaens2023}
{Christiaens}, V., {Gonzalez}, C., {Farkas}, R., {et~al.} 2023, The Journal of
  Open Source Software, 8, 4774

\bibitem[{{Christiaens} {et~al.}(2024){Christiaens}, {Samland}, {Gasman},
  {Temmink}, \& {Perotti}}]{Christiaens2024}
{Christiaens}, V., {Samland}, M., {Gasman}, D., {Temmink}, M., \& {Perotti}, G.
  2024, {MINDS: Hybrid pipeline for the reduction of JWST/MIRI-MRS data},
  Astrophysics Source Code Library, record ascl:2403.007

\bibitem[{{Cieza} {et~al.}(2016){Cieza}, {Casassus}, {Tobin}, {Bos},
  {Williams}, {Perez}, {Zhu}, {Caceres}, {Canovas}, {Dunham}, {Hales},
  {Prieto}, {Principe}, {Schreiber}, {Ruiz-Rodriguez}, \& {Zurlo}}]{cieza2016}
{Cieza}, L.~A., {Casassus}, S., {Tobin}, J., {et~al.} 2016, \nat, 535, 258

\bibitem[{{Cleeves} {et~al.}(2017){Cleeves}, {Bergin}, {{\"O}berg}, {Andrews},
  {Wilner}, \& {Loomis}}]{cleeves2017}
{Cleeves}, L.~I., {Bergin}, E.~A., {{\"O}berg}, K.~I., {et~al.} 2017, \apjl,
  843, L3

\bibitem[{{Cody} {et~al.}(2022){Cody}, {Hillenbrand}, \& {Rebull}}]{cody2022}
{Cody}, A.~M., {Hillenbrand}, L.~A., \& {Rebull}, L.~M. 2022, \aj, 163, 212

\bibitem[{{Colmenares} {et~al.}(2024){Colmenares}, {Bergin}, {Salyk},
  {Pontoppidan}, {Arulanantham}, {Calahan}, {Banzatti}, {Andrews}, {Blake},
  {Ciesla}, {Green}, {Long}, {Lambrechts}, {Najita}, {Pascucci}, {Pinilla},
  {Krijt}, {Trapman}, \& {Jdiscs Collaboration}}]{colmenares2024}
{Colmenares}, M.~J., {Bergin}, E.~A., {Salyk}, C., {et~al.} 2024, \apj, 977,
  173

\bibitem[{{Cruz-S{\'a}enz de Miera} {et~al.}(2023){Cruz-S{\'a}enz de Miera},
  {K{\'o}sp{\'a}l}, {Abrah{\'a}m}, {Claes}, {Manara}, {Wendeborn},
  {Fiorellino}, {Giannini}, {Nisini}, {Sicilia-Aguilar}, {Campbell-White},
  {Alcal{\'a}}, {Banzatti}, {Szab{\'o}}, {Lykou}, {Antoniucci}, {Varga},
  {Siwak}, {Park}, {Nagy}, \& {Kun}}]{cruzsaenzdemiera2023}
{Cruz-S{\'a}enz de Miera}, F., {K{\'o}sp{\'a}l}, {\'A}., {Abrah{\'a}m}, P.,
  {et~al.} 2023, \aap, 678, A88

\bibitem[{{Cushing} {et~al.}(2004){Cushing}, {Vacca}, \&
  {Rayner}}]{cushing2004}
{Cushing}, M.~C., {Vacca}, W.~D., \& {Rayner}, J.~T. 2004, \pasp, 116, 362

\bibitem[{{Cutri} {et~al.}(2003){Cutri}, {Skrutskie}, {van Dyk}, {Beichman},
  {Carpenter}, {Chester}, {Cambresy}, {Evans}, {Fowler}, {Gizis}, {Howard},
  {Huchra}, {Jarrett}, {Kopan}, {Kirkpatrick}, {Light}, {Marsh}, {McCallon},
  {Schneider}, {Stiening}, {Sykes}, {Weinberg}, {Wheaton}, {Wheelock}, \&
  {Zacarias}}]{cutri2003}
{Cutri}, R.~M., {Skrutskie}, M.~F., {van Dyk}, S., {et~al.} 2003, {2MASS All
  Sky Catalog of point sources.} (NASA/IPAC Infrared Science Archive)

\bibitem[{{Czekala} {et~al.}(2016){Czekala}, {Andrews}, {Torres}, {Jensen},
  {Stassun}, {Wilner}, \& {Latham}}]{czekala2016}
{Czekala}, I., {Andrews}, S.~M., {Torres}, G., {et~al.} 2016, \apj, 818, 156

\bibitem[{{Dominik} {et~al.}(2021){Dominik}, {Min}, \& {Tazaki}}]{Dominik2021}
{Dominik}, C., {Min}, M., \& {Tazaki}, R. 2021, {OpTool: Command-line driven
  tool for creating complex dust opacities}, Astrophysics Source Code Library,
  record ascl:2104.010

\bibitem[{{Dorschner} {et~al.}(1995){Dorschner}, {Begemann}, {Henning},
  {J{\"a}ger}, \& {Mutschke}}]{Dorschner1995}
{Dorschner}, J., {Begemann}, B., {Henning}, T., {J{\"a}ger}, C., \& {Mutschke},
  H. 1995, \aap, 300, 503

\bibitem[{{Ercolano} {et~al.}(2021){Ercolano}, {Picogna}, {Monsch}, {Drake}, \&
  {Preibisch}}]{ercolano2021}
{Ercolano}, B., {Picogna}, G., {Monsch}, K., {Drake}, J.~J., \& {Preibisch}, T.
  2021, \mnras, 508, 1675

\bibitem[{{Espaillat} {et~al.}(2023){Espaillat}, {Thanathibodee}, {Pittman},
  {Sturm}, {McClure}, {Calvet}, {Walter}, {Franco-Hern{\'a}ndez}, \& {Muzerolle
  Page}}]{espaillat2023}
{Espaillat}, C.~C., {Thanathibodee}, T., {Pittman}, C.~V., {et~al.} 2023,
  \apjl, 958, L4

\bibitem[{{Favata} {et~al.}(2005){Favata}, {Flaccomio}, {Reale}, {Micela},
  {Sciortino}, {Shang}, {Stassun}, \& {Feigelson}}]{favata2005}
{Favata}, F., {Flaccomio}, E., {Reale}, F., {et~al.} 2005, \apjs, 160, 469

\bibitem[{{Fiorellino} {et~al.}(2022){Fiorellino}, {Park}, {K{\'o}sp{\'a}l}, \&
  {{\'A}brah{\'a}m}}]{fiorellino2022}
{Fiorellino}, E., {Park}, S., {K{\'o}sp{\'a}l}, {\'A}., \& {{\'A}brah{\'a}m},
  P. 2022, \apj, 928, 81

\bibitem[{{Fischer} {et~al.}(2023){Fischer}, {Hillenbrand}, {Herczeg},
  {Johnstone}, {Kospal}, \& {Dunham}}]{fischer2023}
{Fischer}, W.~J., {Hillenbrand}, L.~A., {Herczeg}, G.~J., {et~al.} 2023, in
  Astronomical Society of the Pacific Conference Series, Vol. 534, Protostars
  and Planets VII, ed. S.~{Inutsuka}, Y.~{Aikawa}, T.~{Muto}, K.~{Tomida}, \&
  M.~{Tamura}, 355

\bibitem[{{Flaherty} \& {Muzerolle}(2010)}]{flaherty2010}
{Flaherty}, K.~M. \& {Muzerolle}, J. 2010, \apj, 719, 1733

\bibitem[{{Fraser} {et~al.}(2001){Fraser}, {Collings}, {McCoustra}, \&
  {Williams}}]{fraser2001}
{Fraser}, H.~J., {Collings}, M.~P., {McCoustra}, M. R.~S., \& {Williams}, D.~A.
  2001, \mnras, 327, 1165

\bibitem[{{Gehrels} {et~al.}(2004){Gehrels}, {Chincarini}, {Giommi}, {Mason},
  {Nousek}, {Wells}, {White}, {Barthelmy}, {Burrows}, {Cominsky}, {Hurley},
  {Marshall}, {M{\'e}sz{\'a}ros}, {Roming}, {Angelini}, {Barbier}, {Belloni},
  {Campana}, {Caraveo}, {Chester}, {Citterio}, {Cline}, {Cropper}, {Cummings},
  {Dean}, {Feigelson}, {Fenimore}, {Frail}, {Fruchter}, {Garmire}, {Gendreau},
  {Ghisellini}, {Greiner}, {Hill}, {Hunsberger}, {Krimm}, {Kulkarni}, {Kumar},
  {Lebrun}, {Lloyd-Ronning}, {Markwardt}, {Mattson}, {Mushotzky}, {Norris},
  {Osborne}, {Paczynski}, {Palmer}, {Park}, {Parsons}, {Paul}, {Rees},
  {Reynolds}, {Rhoads}, {Sasseen}, {Schaefer}, {Short}, {Smale}, {Smith},
  {Stella}, {Tagliaferri}, {Takahashi}, {Tashiro}, {Townsley}, {Tueller},
  {Turner}, {Vietri}, {Voges}, {Ward}, {Willingale}, {Zerbi}, \&
  {Zhang}}]{gehrels2004}
{Gehrels}, N., {Chincarini}, G., {Giommi}, P., {et~al.} 2004, \apj, 611, 1005

\bibitem[{{Getman} {et~al.}(2022){Getman}, {Akimkin}, {Arulanantham},
  {K{\'o}sp{\'a}l}, {Semenov}, {Smirnov-Pinchukov}, \& {van
  Terwisga}}]{getman2022}
{Getman}, K.~V., {Akimkin}, V.~V., {Arulanantham}, N., {et~al.} 2022, Research
  Notes of the American Astronomical Society, 6, 64

\bibitem[{{Getman} {et~al.}(2011){Getman}, {Broos}, {Salter}, {Garmire}, \&
  {Hogerheijde}}]{getman2011}
{Getman}, K.~V., {Broos}, P.~S., {Salter}, D.~M., {Garmire}, G.~P., \&
  {Hogerheijde}, M.~R. 2011, \apj, 730, 6

\bibitem[{{Getman} \& {Feigelson}(2021)}]{getman2021}
{Getman}, K.~V. \& {Feigelson}, E.~D. 2021, \apj, 916, 32

\bibitem[{{Getman} {et~al.}(2008{\natexlab{a}}){Getman}, {Feigelson}, {Broos},
  {Micela}, \& {Garmire}}]{getman2008a}
{Getman}, K.~V., {Feigelson}, E.~D., {Broos}, P.~S., {Micela}, G., \&
  {Garmire}, G.~P. 2008{\natexlab{a}}, \apj, 688, 418

\bibitem[{{Getman} {et~al.}(2008{\natexlab{b}}){Getman}, {Feigelson}, {Micela},
  {Jardine}, {Gregory}, \& {Garmire}}]{getman2008b}
{Getman}, K.~V., {Feigelson}, E.~D., {Micela}, G., {et~al.} 2008{\natexlab{b}},
  \apj, 688, 437

\bibitem[{{Getman} {et~al.}(2023){Getman}, {K{\'o}sp{\'a}l}, {Arulanantham},
  {Semenov}, {Smirnov-Pinchukov}, \& {van Terwisga}}]{getman2023}
{Getman}, K.~V., {K{\'o}sp{\'a}l}, {\'A}., {Arulanantham}, N., {et~al.} 2023,
  \apj, 959, 98

\bibitem[{{Glassgold} {et~al.}(2000){Glassgold}, {Feigelson}, \&
  {Montmerle}}]{glassgold2000}
{Glassgold}, A.~E., {Feigelson}, E.~D., \& {Montmerle}, T. 2000, in Protostars
  and Planets IV, ed. V.~{Mannings}, A.~P. {Boss}, \& S.~S. {Russell}, 429

\bibitem[{{Glassgold} {et~al.}(2007){Glassgold}, {Najita}, \&
  {Igea}}]{glassgold2007}
{Glassgold}, A.~E., {Najita}, J.~R., \& {Igea}, J. 2007, \apj, 656, 515

\bibitem[{{Gomez Gonzalez} {et~al.}(2017){Gomez Gonzalez}, {Wertz}, {Absil},
  {Christiaens}, {Defr{\`e}re}, {Mawet}, {Milli}, {Absil}, {Van Droogenbroeck},
  {Cantalloube}, {Hinz}, {Skemer}, {Karlsson}, \& {Surdej}}]{GomezGonzales2017}
{Gomez Gonzalez}, C.~A., {Wertz}, O., {Absil}, O., {et~al.} 2017, \aj, 154, 7

\bibitem[{{G{\"u}nther} \& {Kley}(2002)}]{gunther2002}
{G{\"u}nther}, R. \& {Kley}, W. 2002, \aap, 387, 550

\bibitem[{{Hauschildt} {et~al.}(1999){Hauschildt}, {Allard}, \&
  {Baron}}]{hauschildt1999}
{Hauschildt}, P.~H., {Allard}, F., \& {Baron}, E. 1999, \apj, 512, 377

\bibitem[{{Hawley} {et~al.}(2014){Hawley}, {Davenport}, {Kowalski},
  {Wisniewski}, {Hebb}, {Deitrick}, \& {Hilton}}]{hawley2014}
{Hawley}, S.~L., {Davenport}, J. R.~A., {Kowalski}, A.~F., {et~al.} 2014, \apj,
  797, 121

\bibitem[{{Henden} {et~al.}(2015){Henden}, {Levine}, {Terrell}, \&
  {Welch}}]{henden2015}
{Henden}, A.~A., {Levine}, S., {Terrell}, D., \& {Welch}, D.~L. 2015, in
  American Astronomical Society Meeting Abstracts, Vol. 225, American
  Astronomical Society Meeting Abstracts \#225, 336.16

\bibitem[{{Houge} \& {Krijt}(2023)}]{houge2023}
{Houge}, A. \& {Krijt}, S. 2023, \mnras, 521, 5826

\bibitem[{{Isella} \& {Natta}(2005)}]{isella2005}
{Isella}, A. \& {Natta}, A. 2005, \aap, 438, 899

\bibitem[{{J{\"a}ger} {et~al.}(2003){J{\"a}ger}, {Dorschner}, {Mutschke},
  {Posch}, \& {Henning}}]{Jager2003}
{J{\"a}ger}, C., {Dorschner}, J., {Mutschke}, H., {Posch}, T., \& {Henning}, T.
  2003, \aap, 408, 193

\bibitem[{{J{\"a}ger} {et~al.}(1998){J{\"a}ger}, {Molster}, {Dorschner},
  {Henning}, {Mutschke}, \& {Waters}}]{Jager1998}
{J{\"a}ger}, C., {Molster}, F.~J., {Dorschner}, J., {et~al.} 1998, \aap, 339,
  904

\bibitem[{{Joy}(1945)}]{joy1945}
{Joy}, A.~H. 1945, \apj, 102, 168

\bibitem[{{Kitamura} {et~al.}(2007){Kitamura}, {Pilon}, \&
  {Jonasz}}]{Kitamura2007}
{Kitamura}, R., {Pilon}, L., \& {Jonasz}, M. 2007, \ao, 46, 8118

\bibitem[{{Kobayashi} {et~al.}(2011){Kobayashi}, {Kimura}, {Watanabe},
  {Yamamoto}, \& {M{\"u}ller}}]{kobayashi2011}
{Kobayashi}, H., {Kimura}, H., {Watanabe}, S.-i., {Yamamoto}, T., \&
  {M{\"u}ller}, S. 2011, Earth, Planets and Space, 63, 1067

\bibitem[{{K{\'o}sp{\'a}l} {et~al.}(2012){K{\'o}sp{\'a}l}, {{\'A}brah{\'a}m},
  {Acosta-Pulido}, {Dullemond}, {Henning}, {Kun}, {Leinert}, {Mo{\'o}r}, \&
  {Turner}}]{kospal2012}
{K{\'o}sp{\'a}l}, {\'A}., {{\'A}brah{\'a}m}, P., {Acosta-Pulido}, J.~A.,
  {et~al.} 2012, \apjs, 201, 11

\bibitem[{{K{\'o}sp{\'a}l} {et~al.}(2023){K{\'o}sp{\'a}l}, {{\'A}brah{\'a}m},
  {Diehl}, {Banzatti}, {Bouwman}, {Chen}, {Cruz-S{\'a}enz de Miera}, {Green},
  {Henning}, \& {Rab}}]{kospal2023}
{K{\'o}sp{\'a}l}, {\'A}., {{\'A}brah{\'a}m}, P., {Diehl}, L., {et~al.} 2023,
  \apjl, 945, L7

\bibitem[{{K{\'o}sp{\'a}l} {et~al.}(2018){K{\'o}sp{\'a}l}, {{\'A}brah{\'a}m},
  {Zsidi}, {Vida}, {Szab{\'o}}, {Mo{\'o}r}, \& {P{\'a}l}}]{kospal2018}
{K{\'o}sp{\'a}l}, {\'A}., {{\'A}brah{\'a}m}, P., {Zsidi}, G., {et~al.} 2018,
  \apj, 862, 44

\bibitem[{{Law} {et~al.}(2025){Law}, {Argyriou}, {Gordon}, {Sloan}, {Gasman},
  {Glasse}, {Larson}, {Fletcher}, {Labiano}, \& {Noriega-Crespo}}]{law2025}
{Law}, D.~R., {Argyriou}, I., {Gordon}, K.~D., {et~al.} 2025, \aj, 169, 67

\bibitem[{{Laznevoi} {et~al.}(2025){Laznevoi}, {Akimkin}, {Pavlyuchenkov},
  {Il'in}, {K{\'o}sp{\'a}l}, \& {{\'A}brah{\'a}m}}]{laznevoi2025}
{Laznevoi}, S.~I., {Akimkin}, V.~V., {Pavlyuchenkov}, Y.~N., {et~al.} 2025,
  arXiv e-prints, arXiv:2508.04686

\bibitem[{{Lee} {et~al.}(2019){Lee}, {Lee}, {Baek}, {Aikawa}, {Cieza}, {Yoon},
  {Herczeg}, {Johnstone}, \& {Casassus}}]{lee2019}
{Lee}, J.-E., {Lee}, S., {Baek}, G., {et~al.} 2019, Nature Astronomy, 3, 314

\bibitem[{{Long} {et~al.}(2019){Long}, {Herczeg}, {Harsono}, {Pinilla},
  {Tazzari}, {Manara}, {Pascucci}, {Cabrit}, {Nisini}, {Johnstone}, {Edwards},
  {Salyk}, {Menard}, {Lodato}, {Boehler}, {Mace}, {Liu}, {Mulders}, {Hendler},
  {Ragusa}, {Fischer}, {Banzatti}, {Rigliaco}, {van de Plas}, {Dipierro},
  {Gully-Santiago}, \& {Lopez-Valdivia}}]{long2019}
{Long}, F., {Herczeg}, G.~J., {Harsono}, D., {et~al.} 2019, \apj, 882, 49

\bibitem[{{Menzel} {et~al.}(2023){Menzel}, {Davis}, {Parrish}, {Lawrence},
  {Stewart}, {Cooper}, {Irish}, {Mosier}, {Levine}, {Pitman}, {Walsh},
  {Maghami}, {Thomson}, {Wooldridge}, {Boucarut}, {Feinberg}, {Turner},
  {Kalia}, \& {Bowers}}]{jwst}
{Menzel}, M., {Davis}, M., {Parrish}, K., {et~al.} 2023, \pasp, 135, 058002

\bibitem[{{Min} {et~al.}(2005){Min}, {Hovenier}, \& {de Koter}}]{Min2005}
{Min}, M., {Hovenier}, J.~W., \& {de Koter}, A. 2005, \aap, 432, 909

\bibitem[{{Molyarova} {et~al.}(2018){Molyarova}, {Akimkin}, {Semenov},
  {{\'A}brah{\'a}m}, {Henning}, {K{\'o}sp{\'a}l}, {Vorobyov}, \&
  {Wiebe}}]{molyarova2018}
{Molyarova}, T., {Akimkin}, V., {Semenov}, D., {et~al.} 2018, \apj, 866, 46

\bibitem[{{Mosoni} {et~al.}(2013){Mosoni}, {Sipos}, {{\'A}brah{\'a}m},
  {Mo{\'o}r}, {K{\'o}sp{\'a}l}, {Henning}, {Juh{\'a}sz}, {Kun}, {Leinert},
  {Quanz}, {Ratzka}, {Schegerer}, {van Boekel}, \& {Wolf}}]{mosoni2013}
{Mosoni}, L., {Sipos}, N., {{\'A}brah{\'a}m}, P., {et~al.} 2013, \aap, 552, A62

\bibitem[{{Muzerolle} {et~al.}(2003){Muzerolle}, {Calvet}, {Hartmann}, \&
  {D'Alessio}}]{muzerolle2003}
{Muzerolle}, J., {Calvet}, N., {Hartmann}, L., \& {D'Alessio}, P. 2003, \apjl,
  597, L149

\bibitem[{{Muzerolle} {et~al.}(2019){Muzerolle}, {Flaherty}, {Balog}, {Beck},
  \& {Gutermuth}}]{muzerolle2019}
{Muzerolle}, J., {Flaherty}, K., {Balog}, Z., {Beck}, T., \& {Gutermuth}, R.
  2019, \apj, 877, 29

\bibitem[{{Natta} {et~al.}(2001){Natta}, {Prusti}, {Neri}, {Wooden}, {Grinin},
  \& {Mannings}}]{natta2001}
{Natta}, A., {Prusti}, T., {Neri}, R., {et~al.} 2001, \aap, 371, 186

\bibitem[{{Owen}(2019)}]{owen2019}
{Owen}, J.~E. 2019, Annual Review of Earth and Planetary Sciences, 47, 67

\bibitem[{{Pollack} {et~al.}(1994){Pollack}, {Hollenbach}, {Beckwith},
  {Simonelli}, {Roush}, \& {Fong}}]{pollack1994}
{Pollack}, J.~B., {Hollenbach}, D., {Beckwith}, S., {et~al.} 1994, \apj, 421,
  615

\bibitem[{{Pontoppidan} {et~al.}(2024){Pontoppidan}, {Salyk}, {Banzatti},
  {Zhang}, {Pascucci}, {{\"O}berg}, {Long}, {Romero-Mirza}, {Carr}, {Najita},
  {Blake}, {Arulanantham}, {Andrews}, {Ballering}, {Bergin}, {Calahan}, {Cobb},
  {Colmenares}, {Dickson-Vandervelde}, {Dignan}, {Green}, {Heretz}, {Herczeg},
  {Kalyaan}, {Krijt}, {Pauly}, {Pinilla}, {Trapman}, \&
  {Xie}}]{pontoppidan2024}
{Pontoppidan}, K.~M., {Salyk}, C., {Banzatti}, A., {et~al.} 2024, \apj, 963,
  158

\bibitem[{{Pouilly} {et~al.}(2023){Pouilly}, {Kochukhov}, {K{\'o}sp{\'a}l},
  {Hahlin}, {Carmona}, \& {{\'A}brah{\'a}m}}]{pouilly2023}
{Pouilly}, K., {Kochukhov}, O., {K{\'o}sp{\'a}l}, {\'A}., {et~al.} 2023,
  \mnras, 518, 5072

\bibitem[{{Rab} {et~al.}(2017{\natexlab{a}}){Rab}, {Elbakyan}, {Vorobyov},
  {G{\"u}del}, {Dionatos}, {Audard}, {Kamp}, {Thi}, {Woitke}, \&
  {Postel}}]{rab2017a}
{Rab}, C., {Elbakyan}, V., {Vorobyov}, E., {et~al.} 2017{\natexlab{a}}, \aap,
  604, A15

\bibitem[{{Rab} {et~al.}(2017{\natexlab{b}}){Rab}, {G{\"u}del}, {Padovani},
  {Kamp}, {Thi}, {Woitke}, \& {Aresu}}]{rab2017b}
{Rab}, C., {G{\"u}del}, M., {Padovani}, M., {et~al.} 2017{\natexlab{b}}, \aap,
  603, A96

\bibitem[{{Rayner} {et~al.}(2003){Rayner}, {Toomey}, {Onaka}, {Denault},
  {Stahlberger}, {Vacca}, {Cushing}, \& {Wang}}]{rayner2003}
{Rayner}, J.~T., {Toomey}, D.~W., {Onaka}, P.~M., {et~al.} 2003, \pasp, 115,
  362

\bibitem[{{Salter} {et~al.}(2010){Salter}, {K{\'o}sp{\'a}l}, {Getman},
  {Hogerheijde}, {van Kempen}, {Carpenter}, {Blake}, \& {Wilner}}]{salter2010}
{Salter}, D.~M., {K{\'o}sp{\'a}l}, {\'A}., {Getman}, K.~V., {et~al.} 2010,
  \aap, 521, A32

\bibitem[{{Salyk}(2020)}]{salyk2020}
{Salyk}, C. 2020, {slabspec: Python code for producing LTE slab model molecular
  spectra}

\bibitem[{{Smith} {et~al.}(2025){Smith}, {Romero-Mirza}, {Banzatti}, {Rab},
  {Abraham}, {Kospal}, {Claes}, {Manara}, {Oberg}, {Bouwman}, {Cruz-Saenz de
  Miera}, \& {Green}}]{smith2025}
{Smith}, S.~A., {Romero-Mirza}, C.~E., {Banzatti}, A., {et~al.} 2025, arXiv
  e-prints, arXiv:2504.13377

\bibitem[{{Suto} {et~al.}(2006){Suto}, {Sogawa}, {Tachibana}, {Koike},
  {Karoji}, {Tsuchiyama}, {Chihara}, {Mizutani}, {Akedo}, {Ogiso}, {Fukui}, \&
  {Ohara}}]{Suto2006}
{Suto}, H., {Sogawa}, H., {Tachibana}, S., {et~al.} 2006, \mnras, 370, 1599

\bibitem[{{Sytov} {et~al.}(2011){Sytov}, {Kaigorodov}, {Fateeva}, \&
  {Bisikalo}}]{sytov2011}
{Sytov}, A.~Y., {Kaigorodov}, P.~V., {Fateeva}, A.~M., \& {Bisikalo}, D.~V.
  2011, Astronomy Reports, 55, 793

\bibitem[{{Takami} {et~al.}(2018){Takami}, {Fu}, {Liu}, {Karr}, {Hashimoto},
  {Kudo}, {Vorobyov}, {K{\'o}sp{\'a}l}, {Scicluna}, {Dong}, {Tamura}, {Pyo},
  {Fukagawa}, {Tsuribe}, {Dunham}, {Henning}, \& {de Leon}}]{takami2018}
{Takami}, M., {Fu}, G., {Liu}, H.~B., {et~al.} 2018, \apj, 864, 20

\bibitem[{{Temmink} {et~al.}(2024){Temmink}, {van Dishoeck}, {Grant}, {Tabone},
  {Gasman}, {Christiaens}, {Samland}, {Argyriou}, {Perotti}, {G{\"u}del},
  {Henning}, {Lagage}, {Abergel}, {Absil}, {Barrado}, {Caratti o Garatti},
  {Glauser}, {Kamp}, {Lahuis}, {Olofsson}, {Ray}, {Scheithauer},
  {Vandenbussche}, {Waters}, {Arabhavi}, {Jang}, {Kanwar},
  {Morales-Calder{\'o}n}, {Rodgers-Lee}, {Schreiber}, {Schwarz}, \&
  {Colina}}]{Temmink2024}
{Temmink}, M., {van Dishoeck}, E.~F., {Grant}, S.~L., {et~al.} 2024, \aap, 686,
  A117

\bibitem[{{Tofflemire} {et~al.}(2025){Tofflemire}, {Manara}, {Banzatti},
  {Pontoppidan}, {Najita}, {Nisini}, {Whelan}, {Campbell-White}, {Alqubelat},
  {Kraus}, {Rab}, {Houge}, {Krijt}, {Muzerolle}, {Fiorellino}, {Benisty},
  {Tychoniec}, {Salyk}, {Bourdarot}, \& {Hyden}}]{tofflemire2025}
{Tofflemire}, B.~M., {Manara}, C.~F., {Banzatti}, A., {et~al.} 2025, \apj, 985,
  224

\bibitem[{{Topchieva} {et~al.}(2025){Topchieva}, {Molyarova}, \&
  {Vorobyov}}]{topchieva2025}
{Topchieva}, A., {Molyarova}, T., \& {Vorobyov}, E. 2025, arXiv e-prints,
  arXiv:2505.07718

\bibitem[{{Vacca} {et~al.}(2003){Vacca}, {Cushing}, \& {Rayner}}]{vacca2003}
{Vacca}, W.~D., {Cushing}, M.~C., \& {Rayner}, J.~T. 2003, \pasp, 115, 389

\bibitem[{{Visser} {et~al.}(2015){Visser}, {Bergin}, \&
  {J{\o}rgensen}}]{visser2015}
{Visser}, R., {Bergin}, E.~A., \& {J{\o}rgensen}, J.~K. 2015, \aap, 577, A102

\bibitem[{{Visser} {et~al.}(2009){Visser}, {van Dishoeck}, \&
  {Black}}]{visser2009}
{Visser}, R., {van Dishoeck}, E.~F., \& {Black}, J.~H. 2009, \aap, 503, 323

\bibitem[{{Waggoner} {et~al.}(2023){Waggoner}, {Cleeves}, {Loomis}, {Aikawa},
  {Bae}, {Bergner}, {Booth}, {Calahan}, {Cataldi}, {Law}, {Le Gal}, {Long},
  {{\"O}berg}, {Teague}, \& {Wilner}}]{waggoner2023}
{Waggoner}, A.~R., {Cleeves}, L.~I., {Loomis}, R.~A., {et~al.} 2023, \apj, 956,
  103

\bibitem[{{Walsh} {et~al.}(2012){Walsh}, {Nomura}, {Millar}, \&
  {Aikawa}}]{walsh2012}
{Walsh}, C., {Nomura}, H., {Millar}, T.~J., \& {Aikawa}, Y. 2012, \apj, 747,
  114

\bibitem[{{Washinoue} {et~al.}(2024){Washinoue}, {Takasao}, \&
  {Furuya}}]{washinoue2024}
{Washinoue}, H., {Takasao}, S., \& {Furuya}, K. 2024, \apj, 976, 25

\bibitem[{{Woitke} {et~al.}(2024){Woitke}, {Thi}, {Arabhavi}, {Kamp},
  {K{\'o}sp{\'a}l}, \& {{\'A}brah{\'a}m}}]{woitke2024}
{Woitke}, P., {Thi}, W.~F., {Arabhavi}, A.~M., {et~al.} 2024, \aap, 683, A219

\bibitem[{{Wright} {et~al.}(2023){Wright}, {Rieke}, {Glasse}, {Ressler},
  {Garc{\'\i}a Mar{\'\i}n}, {Aguilar}, {Alberts}, {{\'A}lvarez-M{\'a}rquez},
  {Argyriou}, {Banks}, {Baudoz}, {Boccaletti}, {Bouchet}, {Bouwman}, {Brandl},
  {Breda}, {Bright}, {Cale}, {Colina}, {Cossou}, {Coulais}, {Cracraft}, {De
  Meester}, {Dicken}, {Engesser}, {Etxaluze}, {Fox}, {Friedman}, {Fu},
  {Gasman}, {G{\'a}sp{\'a}r}, {Gastaud}, {Geers}, {Glauser}, {Gordon},
  {Greene}, {Greve}, {Grundy}, {G{\"u}del}, {Guillard}, {Haderlein},
  {Hashimoto}, {Henning}, {Hines}, {Holler}, {Detre}, {Jahromi}, {James},
  {Jones}, {Justtanont}, {Kavanagh}, {Kendrew}, {Klaassen}, {Krause},
  {Labiano}, {Lagage}, {Lambros}, {Larson}, {Law}, {Lee}, {Libralato}, {Lorenzo
  Alverez}, {Meixner}, {Morrison}, {Mueller}, {Murray}, {Mycroft}, {Myers},
  {Nayak}, {Naylor}, {Nickson}, {Noriega-Crespo}, {{\"O}stlin}, {O'Sullivan},
  {Ottens}, {Patapis}, {Penanen}, {Pietraszkiewicz}, {Ray}, {Regan},
  {Roteliuk}, {Royer}, {Samara-Ratna}, {Samuelson}, {Sargent}, {Scheithauer},
  {Schneider}, {Schreiber}, {Shaughnessy}, {Sheehan}, {Shivaei}, {Sloan},
  {Tamas}, {Teague}, {Temim}, {Tikkanen}, {Tustain}, {van Dishoeck},
  {Vandenbussche}, {Weilert}, {Whitehouse}, \& {Wolff}}]{miri}
{Wright}, G.~S., {Rieke}, G.~H., {Glasse}, A., {et~al.} 2023, \pasp, 135,
  048003

\bibitem[{{Xie} {et~al.}(2023){Xie}, {Pascucci}, {Long}, {Pontoppidan},
  {Banzatti}, {Kalyaan}, {Salyk}, {Liu}, {Najita}, {Pinilla}, {Arulanantham},
  {Herczeg}, {Carr}, {Bergin}, {Ballering}, {Krijt}, {Blake}, {Zhang},
  {{\"O}berg}, {Green}, \& {Jdiscs Collaboration}}]{xie2023}
{Xie}, C., {Pascucci}, I., {Long}, F., {et~al.} 2023, \apjl, 959, L25

\bibitem[{{Zeidler} {et~al.}(2015){Zeidler}, {Mutschke}, \&
  {Posch}}]{Zeidler2015}
{Zeidler}, S., {Mutschke}, H., \& {Posch}, T. 2015, \apj, 798, 125

\bibitem[{{Zwicky} {et~al.}(2024){Zwicky}, {Molyarova}, {Akimkin},
  {Smirnov-Pinchukov}, {Semenov}, {K{\'o}sp{\'a}l}, \&
  {{\'A}brah{\'a}m}}]{zwicky2024}
{Zwicky}, L., {Molyarova}, T., {Akimkin}, V., {et~al.} 2024, \mnras, 527, 7652

\end{thebibliography}

\begin{appendix}

\onecolumn

\section{Accretion variations}

Table \ref{tab:irtf} lists the observing dates and modified Julian dates for our NASA IRTF/SpeX observations of DQ\,Tau. The table also contains the orbital phase at the time of the observations using orbital parameters from \citet{czekala2016}, the extinction value towards DQ\,Tau ($A_V$), as well as the accretion luminosities as determined by the method summarized in Sec.~\ref{sec:acc}. To calculate accretion rates from the accretion luminosities, we used Equation (4) and stellar parameters from \citet{fiorellino2022}: $\dot{M}_{\rm acc} = 1.25L_{\rm acc}R_*/(GM_*)$ with $R_*=2.58\,$R$_{\odot}$ and $M_*=1.52\,$M$_{\odot}$.

\begin{figure}
\centering
\includegraphics[width=0.6\textwidth]{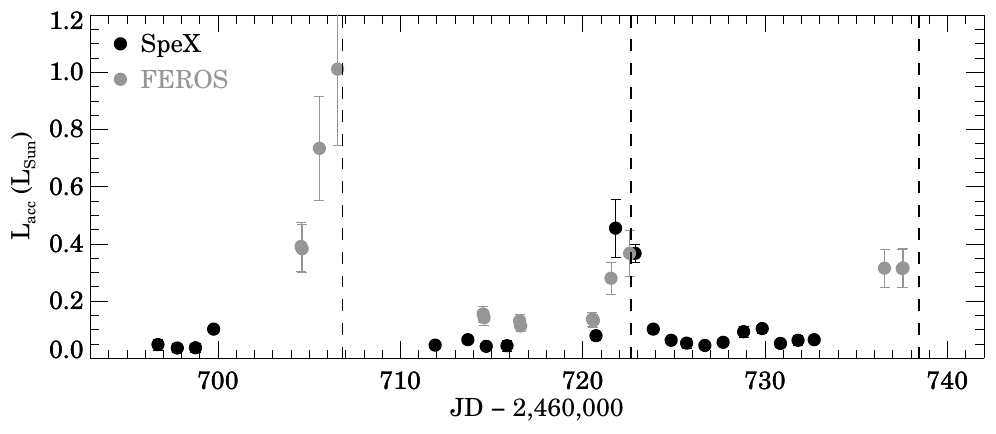}
\includegraphics[width=0.6\textwidth]{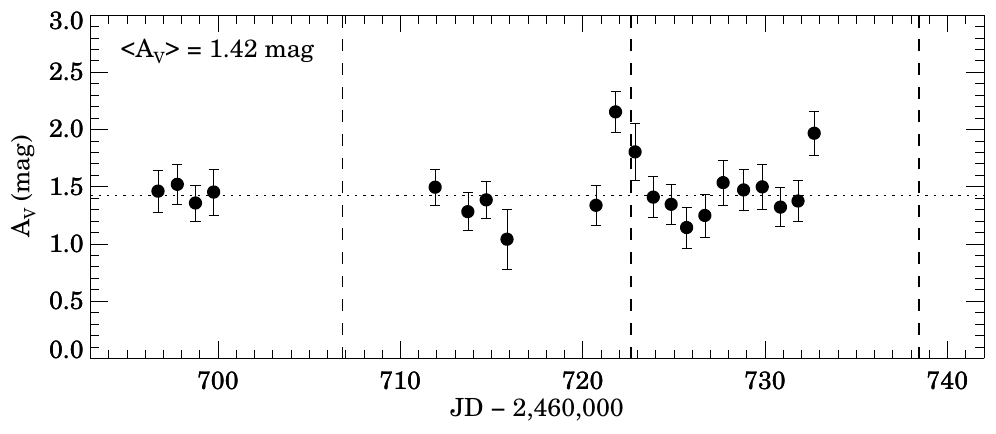}
\caption{Accretion luminosity (top) and extinction ($A_V$, bottom) as a function of time for DQ\,Tau. Vertical dashed lines indicate the epochs of periastrons.}\label{fig:lacc_av}
\end{figure}

\begin{figure}[h!]
\centering
\includegraphics[width=0.49\textwidth]{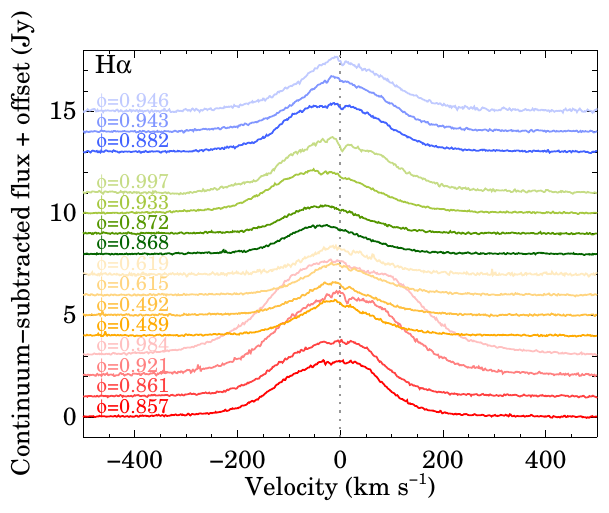}
\caption{Profiles of the H${\alpha}$ line in DQ\,Tau observed with the FEROS spectrograph on the MPG 2.2\,m telescope. For clarity, the spectra were shifted along the y axis by +1, +2, +3, \dots\,Jy. The binary orbital phases at the time of the observations are indicated to the left.}\label{fig:halpha}
\end{figure}

Table \ref{tab:feros} lists the observing dates and modified Julian dates for our MPG\,2.2\,m/FEROS observations of DQ\,Tau. The table also contains the orbital phase at the time of the observations again using orbital parameters from \citet{czekala2016}, as well as the accretion luminosities calculated for a fixed $A_V=1.42\,$mag. Accretion rates were calculated in the same way as for the IRTF-based data. Fig.~\ref{fig:lacc_av} shows the obtained $L_{\rm acc}$ and $A_V$ values plotted as a function of time. Fig.~\ref{fig:halpha} shows the profiles of the H$\alpha$ line from the FEROS spectra.

\begin{table*}[h!]
\caption{Extinction and accretion values for DQ\,Tau from the IRTF/SpeX spectra.}
\label{tab:irtf}
\centering
\begin{tabular}{c c c c c c}
\hline\hline
Date        & MJD      & Phase & $A_V$         & $L_{\rm acc}$ & $\dot{M}_{\rm acc}$     \\
            &          &       & (mag)         & ($L_{\odot}$) & ($M_{\odot}$\,yr$^{-1}$) \\
\hline
2025-Jan-21 & 60696.212 & 0.359 & 1.46$\pm$0.18 & 0.048$\pm$0.020 & 3.21e-09\\
2025-Jan-22 & 60697.270 & 0.426 & 1.52$\pm$0.17 & 0.036$\pm$0.015 & 2.45e-09\\
2025-Jan-23 & 60698.267 & 0.489 & 1.36$\pm$0.16 & 0.037$\pm$0.016 & 2.49e-09\\
2025-Jan-24 & 60699.264 & 0.553 & 1.45$\pm$0.20 & 0.102$\pm$0.009 & 6.91e-09\\
2025-Feb-05 & 60711.411 & 0.321 & 1.50$\pm$0.16 & 0.046$\pm$0.011 & 3.08e-09\\
2025-Feb-07 & 60713.204 & 0.435 & 1.28$\pm$0.17 & 0.065$\pm$0.011 & 4.37e-09\\
2025-Feb-08 & 60714.212 & 0.499 & 1.38$\pm$0.16 & 0.042$\pm$0.009 & 2.82e-09\\
2025-Feb-09 & 60715.352 & 0.571 & 1.04$\pm$0.26 & 0.044$\pm$0.019 & 3.00e-09\\
2025-Feb-14 & 60720.232 & 0.879 & 1.34$\pm$0.17 & 0.079$\pm$0.015 & 5.33e-09\\
2025-Feb-15 & 60721.305 & 0.947 & 2.15$\pm$0.17 & 0.455$\pm$0.101 & 3.07e-08\\
2025-Feb-16 & 60722.375 & 0.015 & 1.80$\pm$0.25 & 0.367$\pm$0.031 & 2.48e-08\\
2025-Feb-17 & 60723.372 & 0.078 & 1.41$\pm$0.18 & 0.102$\pm$0.015 & 6.90e-09\\
2025-Feb-18 & 60724.354 & 0.140 & 1.35$\pm$0.18 & 0.063$\pm$0.012 & 4.24e-09\\
2025-Feb-19 & 60725.200 & 0.194 & 1.14$\pm$0.18 & 0.053$\pm$0.017 & 3.57e-09\\
2025-Feb-20 & 60726.200 & 0.257 & 1.25$\pm$0.19 & 0.045$\pm$0.015 & 3.01e-09\\
2025-Feb-21 & 60727.203 & 0.321 & 1.54$\pm$0.19 & 0.056$\pm$0.011 & 3.76e-09\\
2025-Feb-22 & 60728.326 & 0.392 & 1.47$\pm$0.18 & 0.093$\pm$0.018 & 6.31e-09\\
2025-Feb-23 & 60729.348 & 0.456 & 1.50$\pm$0.20 & 0.104$\pm$0.015 & 7.04e-09\\
2025-Feb-24 & 60730.347 & 0.520 & 1.32$\pm$0.17 & 0.052$\pm$0.012 & 3.51e-09\\
2025-Feb-25 & 60731.311 & 0.581 & 1.37$\pm$0.18 & 0.063$\pm$0.016 & 4.24e-09\\
2025-Feb-26 & 60732.198 & 0.637 & 1.97$\pm$0.19 & 0.065$\pm$0.014 & 4.39e-09\\
\hline
\end{tabular}
\end{table*}

\begin{table*}[h!]
\caption{Accretion values for DQ\,Tau from the MPG\,2.2\,m/FEROS spectra.}
\label{tab:feros}
\centering
\begin{tabular}{c c c c c c}
\hline\hline
Date        & MJD      & Phase & $A_V^{a}$ & $L_{\rm acc}$ & $\dot{M}_{\rm acc}$     \\
            &          &       & (mag)     & ($L_{\odot}$) & ($M_{\odot}$\,yr$^{-1}$) \\
\hline
2025-Jan-29 & 60704.060 & 0.856 & 1.42     & 0.391$\pm$0.086 & 2.64e-08\\
2025-Jan-29 & 60704.116 & 0.860 & 1.42     & 0.383$\pm$0.084 & 2.59e-08\\
2025-Jan-30 & 60705.063 & 0.920 & 1.42     & 0.734$\pm$0.183 & 4.95e-08\\
2025-Jan-31 & 60706.061 & 0.983 & 1.42     & 1.011$\pm$0.266 & 6.83e-08\\

2025-Feb-08 & 60714.040 & 0.488 & 1.42     & 0.155$\pm$0.028 & 1.05e-08\\
2025-Feb-08 & 60714.094 & 0.491 & 1.42     & 0.142$\pm$0.025 & 9.59e-09\\
2025-Feb-10 & 60716.043 & 0.614 & 1.42     & 0.131$\pm$0.023 & 8.88e-09\\
2025-Feb-10 & 60716.102 & 0.618 & 1.42     & 0.113$\pm$0.019 & 7.61e-09\\

2025-Feb-14 & 60720.038 & 0.867 & 1.42     & 0.137$\pm$0.024 & 9.23e-09\\
2025-Feb-14 & 60720.090 & 0.870 & 1.42     & 0.132$\pm$0.023 & 8.93e-09\\
2025-Feb-15 & 60721.056 & 0.932 & 1.42     & 0.280$\pm$0.057 & 1.89e-08\\
2025-Feb-16 & 60722.075 & 0.996 & 1.42     & 0.367$\pm$0.080 & 2.48e-08\\

2025-Mar-02 & 60736.054 & 0.881 & 1.42     & 0.315$\pm$0.066 & 2.13e-08\\
2025-Mar-03 & 60737.023 & 0.942 & 1.42     & 0.314$\pm$0.066 & 2.12e-08\\
2025-Mar-03 & 60737.071 & 0.945 & 1.42     & 0.316$\pm$0.067 & 2.14e-08\\
\hline
\end{tabular}
\tablefoot{$^{a}$ fixed to 1.42\,mag}
\end{table*}

\FloatBarrier

\section{Dust mineralogy in DQ\,Tau}

\begin{figure*}
\centering
\includegraphics[width=0.45\textwidth]{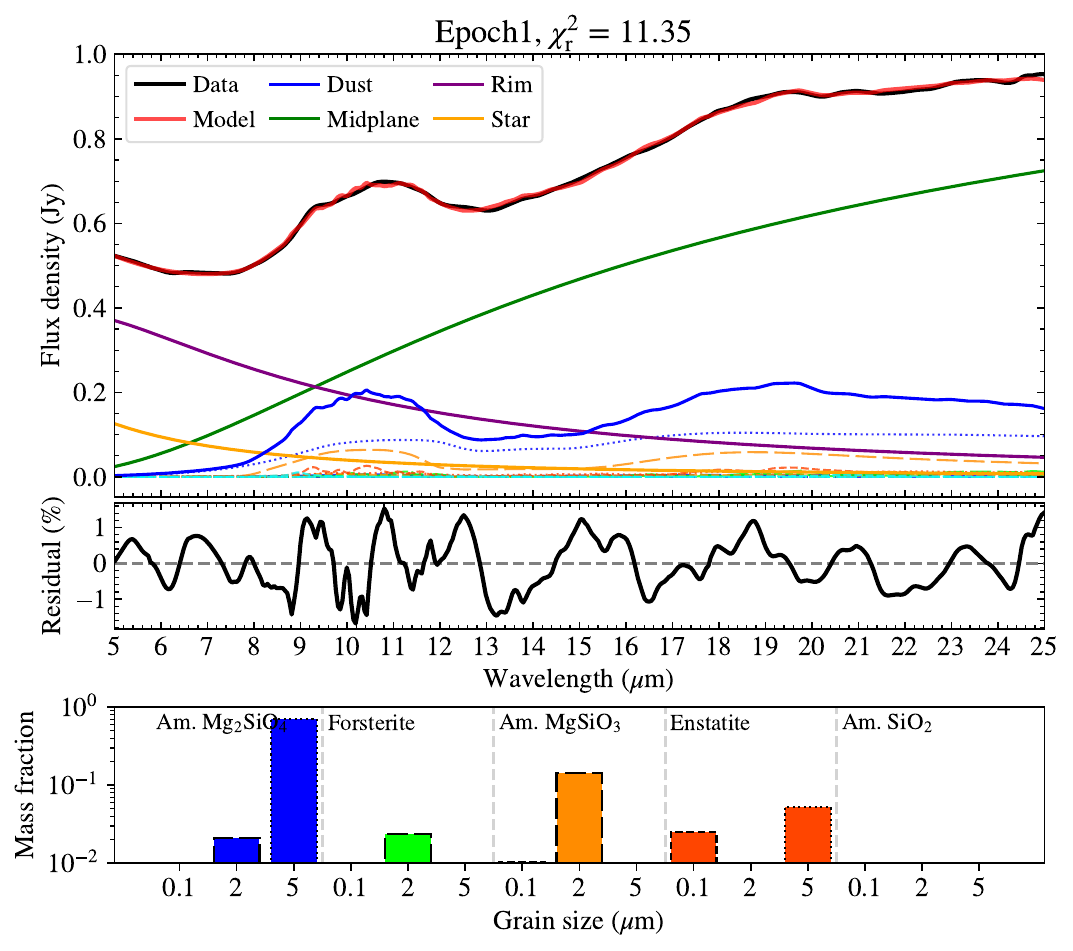}
\includegraphics[width=0.45\textwidth]{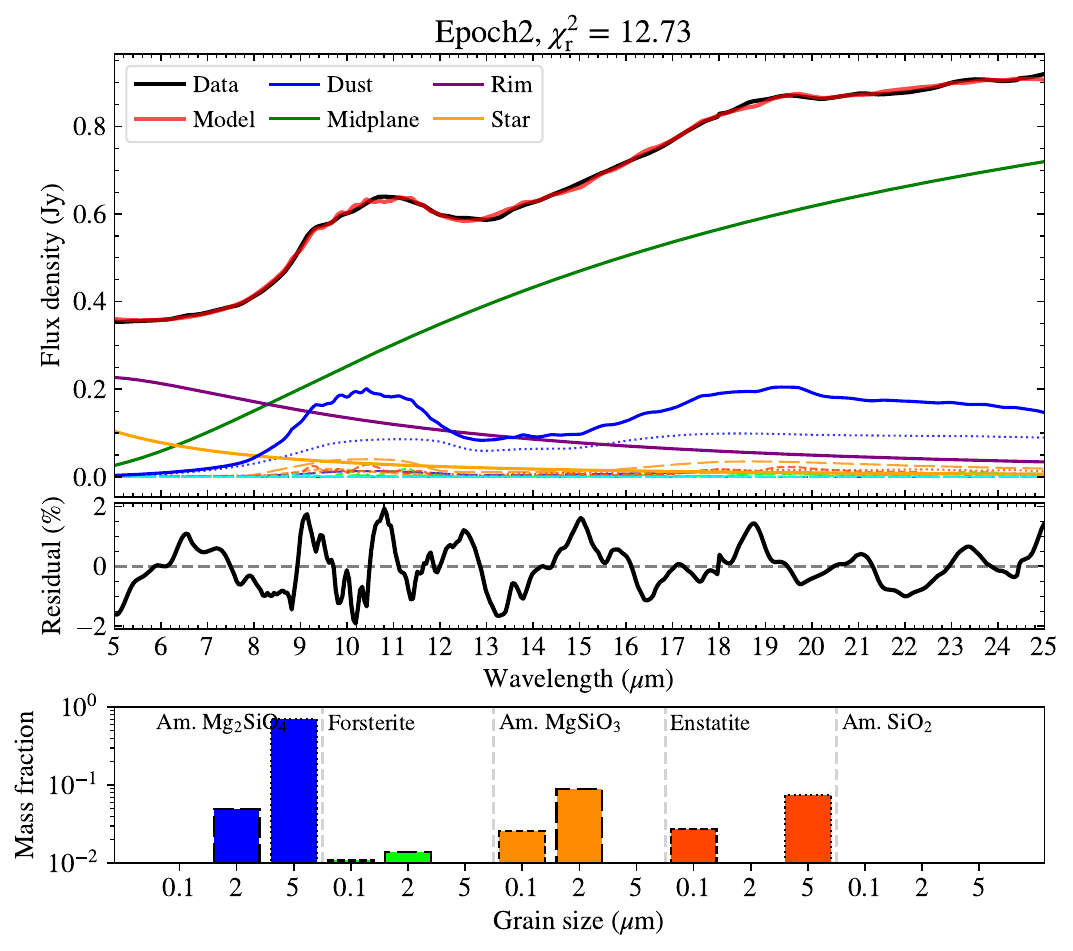}
\includegraphics[width=0.45\textwidth]{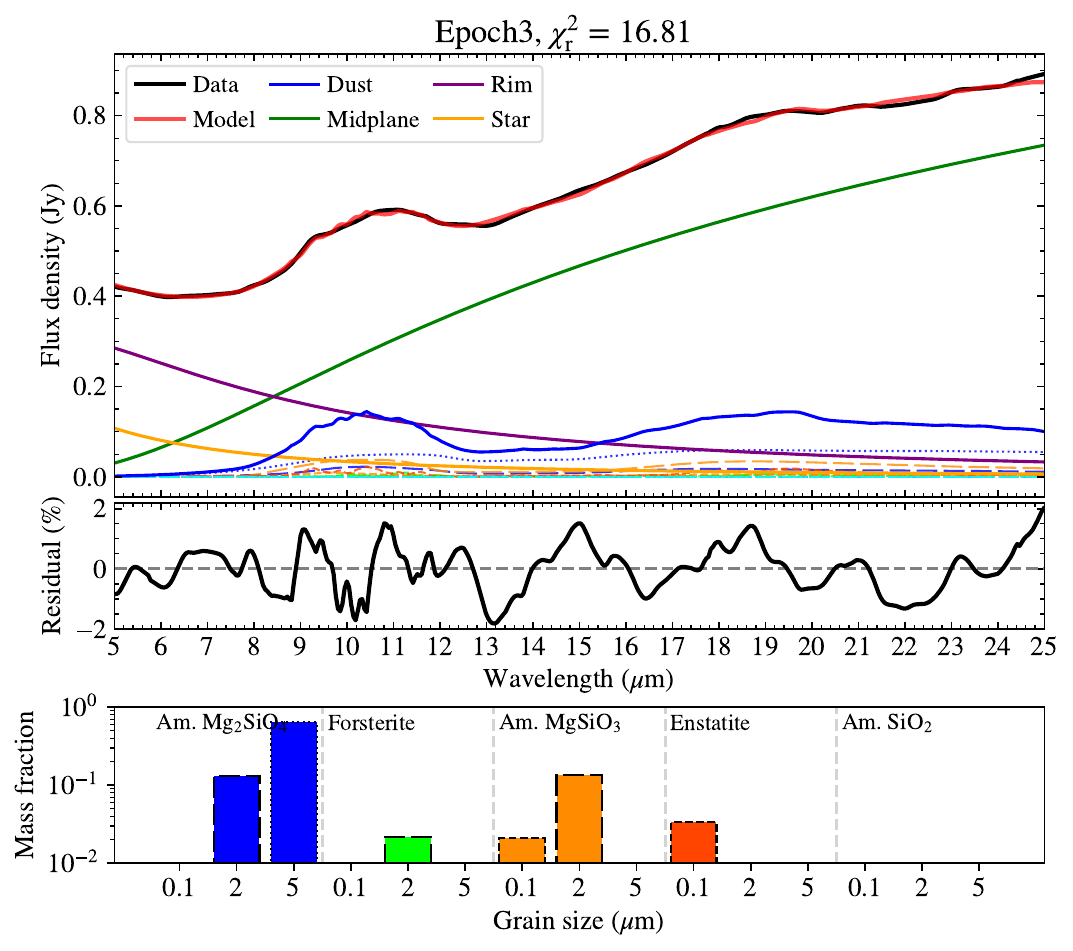}
\includegraphics[width=0.45\textwidth]{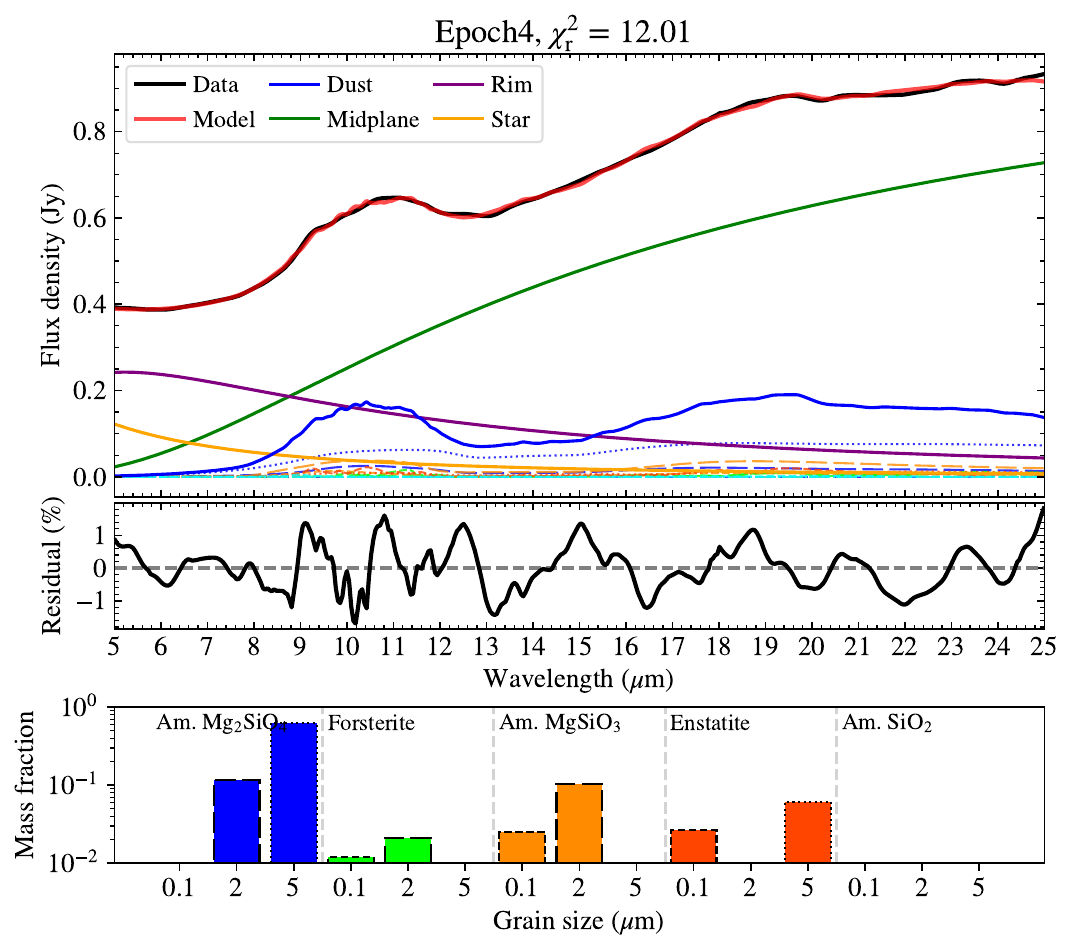}
\caption{Results of the oldDHS dust model fitting. Top: various model components along with the total model (red) and the line-free observed spectrum (black). Middle: residuals. Bottom: Relative mass fractions of the various dust components.}\label{fig:oldDHS}
\end{figure*}

\begin{figure*}
\centering
\includegraphics[width=0.45\textwidth]{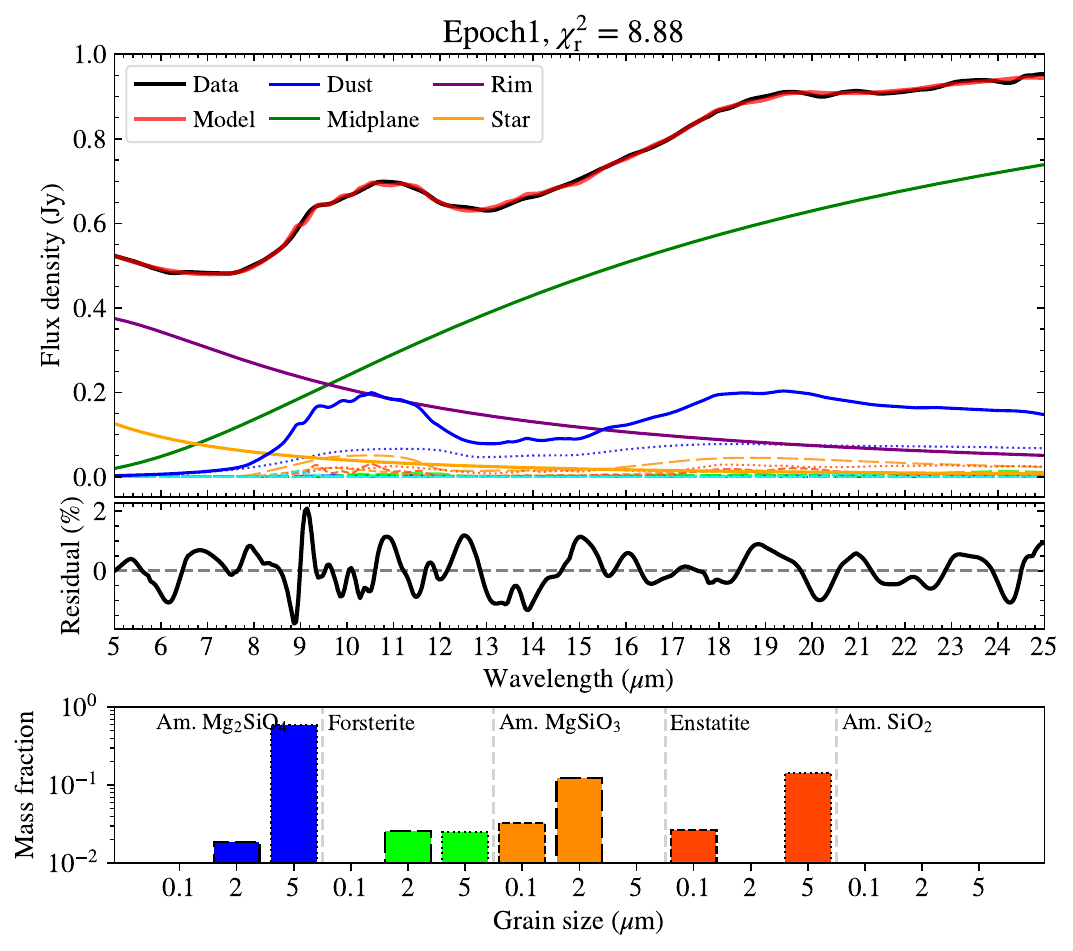}
\includegraphics[width=0.45\textwidth]{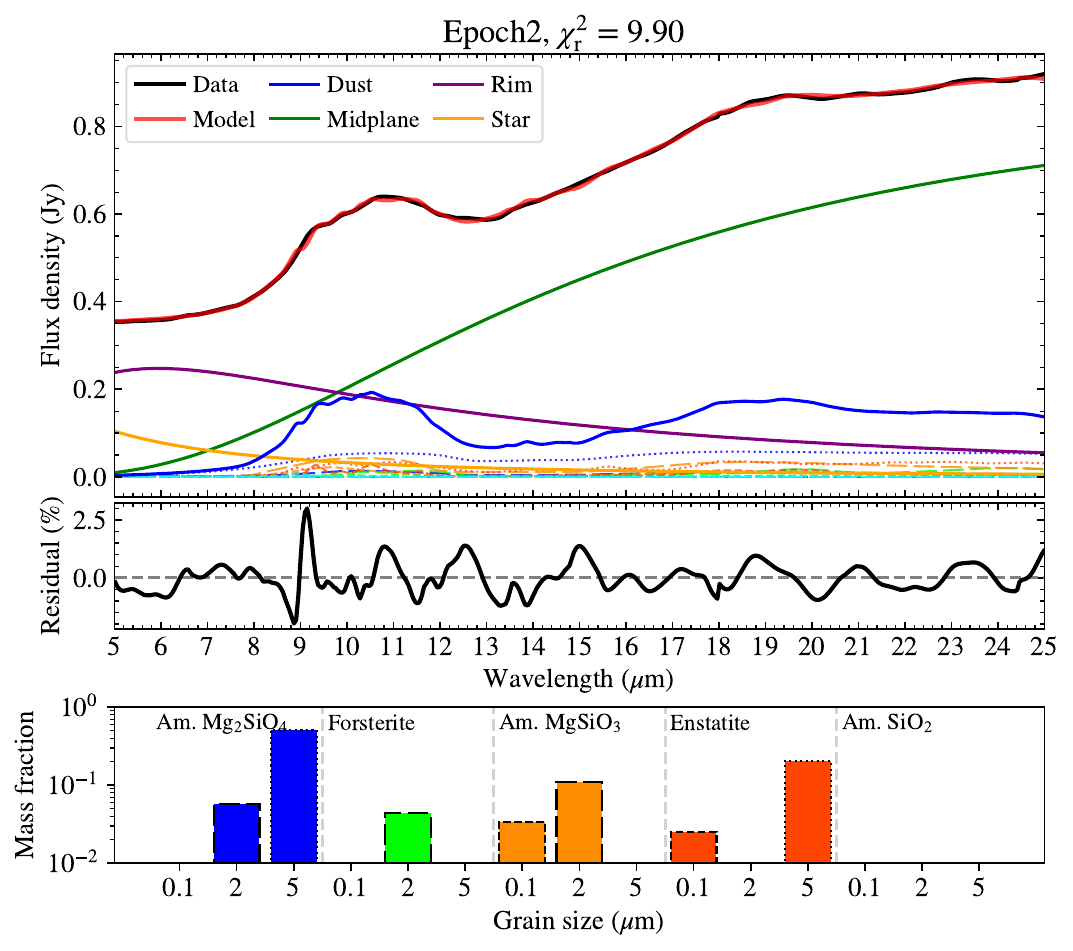}
\includegraphics[width=0.45\textwidth]{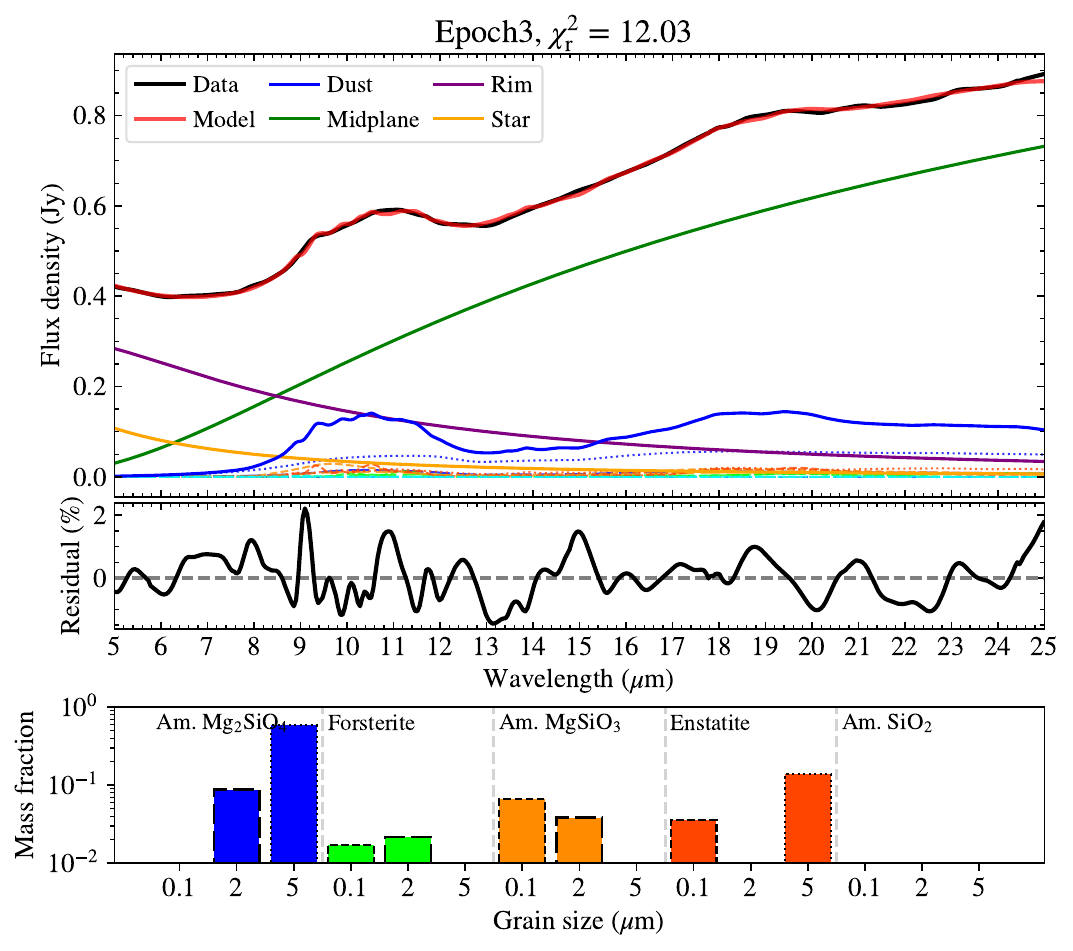}
\includegraphics[width=0.45\textwidth]{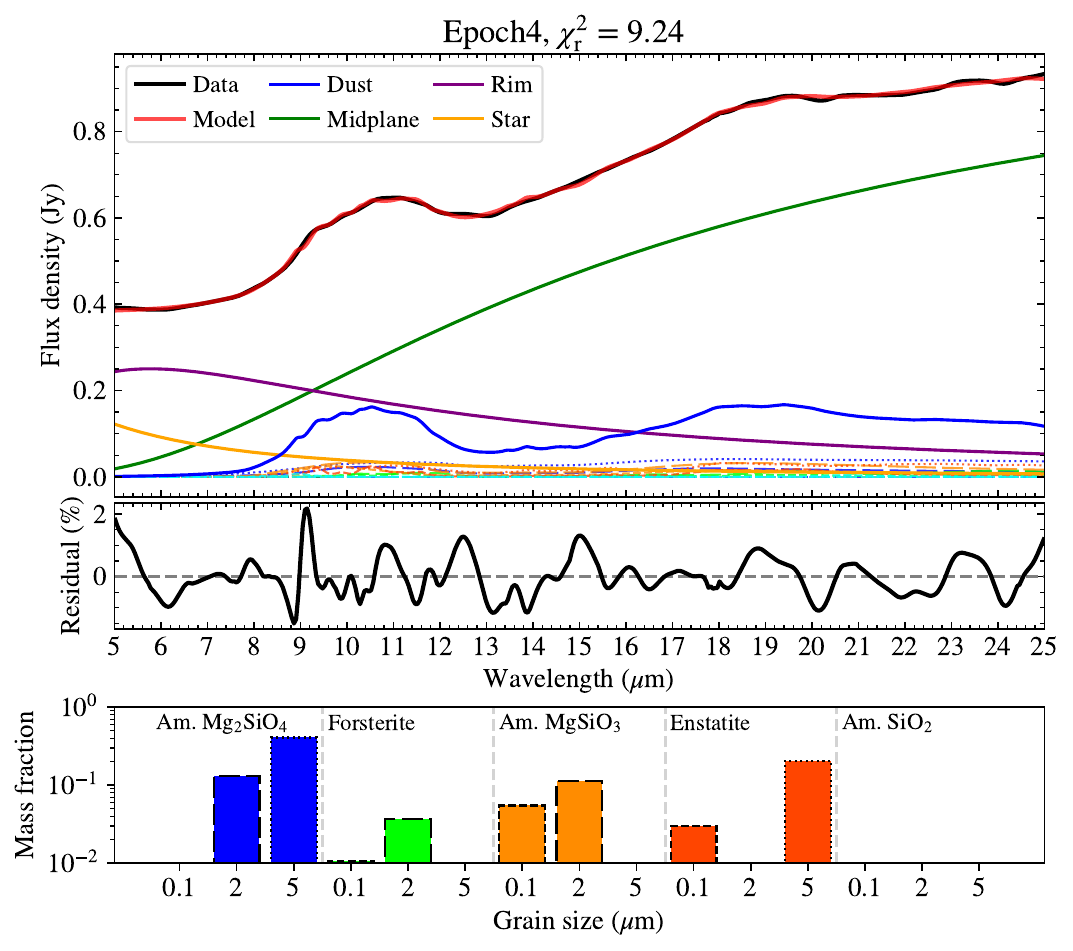}
\caption{Same as Fig.~\ref{fig:oldDHS} but for the newDHS dust model.}\label{fig:newDHS}
\end{figure*}

Figures~\ref{fig:oldDHS}-\ref{fig:newDHS} and Tables \ref{tab:oldDHS}-\ref{tab:newDHS} display the results of our dust mineralogy fitting.

\begin{table*}
\caption{Mass fractions for various dust components of DQ\,Tau in percent for the oldDHS dust model.}\label{tab:oldDHS}
\centering
\begin{tabular}{l c c c c c c c}
\hline\hline
Dust component & Grain size ($\mu$m) & Epoch 1 & Epoch 2 & Epoch 3 & Epoch 4 & Avg & Std \\
\hline
Amorphous olivine & 0.1 & $0.00$ & $0.00$ & $0.00$ & $0.00$ & 0.00 & 0.00\\
Amorphous olivine & 2.0 & $2.12$ & $4.90$ & $13.07$ & $11.62$ & 7.93 & 4.56\\
Amorphous olivine & 5.0 & $71.11$ & $70.47$ & $64.54$ & $62.88$ & 67.25 & 3.59\\
\hline
Crystalline forsterite & 0.1 & $0.77$ & $1.08$ & $0.91$ & $1.20$ & 0.99 & 0.16\\
Crystalline forsterite & 2.0 & $2.39$ & $1.41$ & $2.19$ & $2.13$ & 2.03 & 0.37\\
Crystalline forsterite & 5.0 & $0.00$ & $0.00$ & $0.00$ & $0.00$ & 0.00 & 0.00\\
\hline
Amorphous pyroxene & 0.1 & $1.03$ & $2.56$ & $2.10$ & $2.51$ & 2.05 & 0.62\\
Amorphous pyroxene & 2.0 & $14.30$ & $9.02$ & $13.45$ & $10.49$ & 11.82 & 2.15\\
Amorphous pyroxene & 5.0 & $0.00$ & $0.00$ & $0.00$ & $0.00$ & 0.00 & 0.00\\
\hline
Crystalline enstatite & 0.1 & $2.53$ & $2.72$ & $3.34$ & $2.69$ & 2.82 & 0.31\\
Crystalline enstatite & 2.0 & $0.00$ & $0.00$ & $0.00$ & $0.00$ & 0.00 & 0.00\\
Crystalline enstatite & 5.0 & $5.30$ & $7.50$ & $0.41$ & $6.17$ & 4.85 & 2.68\\
\hline
Amorphous silica & 0.1 & $0.46$ & $0.33$ & $0.00$ & $0.29$ & 0.27 & 0.17\\
Amorphous silica & 2.0 & $0.00$ & $0.00$ & $0.00$ & $0.00$ & 0.00 & 0.00\\
Amorphous silica & 5.0 & $0.00$ & $0.00$ & $0.00$ & $0.00$ & 0.00 & 0.00\\
\hline
\end{tabular}
\end{table*}

\begin{table*}
\caption{Mass fractions for various dust components of DQ\,Tau in percent for the newDHS dust model.}\label{tab:newDHS}
\centering
\begin{tabular}{l c c c c c c c}
\hline\hline
Dust component & Grain size ($\mu$m)  & Epoch 1 & Epoch 2 & Epoch 3 & Epoch 4 & Avg & Std\\
\hline
Amorphous olivine & 0.1 & $0.00$ & $0.00$ & $0.00$ & $0.00$ & 0.00 & 0.00\\
Amorphous olivine & 2.0 & $1.89$ & $5.81$ & $8.82$ & $13.06$ & 7.39 & 4.09\\
Amorphous olivine & 5.0 & $58.89$ & $51.36$ & $59.37$ & $40.98$ & 52.65 & 7.45\\
\hline
Crystalline forsterite & 0.1 & $0.80$ & $0.61$ & $1.70$ & $1.07$ & 1.04 & 0.41\\
Crystalline forsterite & 2.0 & $2.60$ & $4.35$ & $2.16$ & $3.66$ & 3.19 & 0.86\\
Crystalline forsterite & 5.0 & $2.54$ & $0.00$ & $0.00$ & $0.00$ & 0.63 & 1.10\\
\hline
Amorphous pyroxene & 0.1 & $3.31$ & $3.42$ & $6.60$ & $5.50$ & 4.71 & 1.40\\
Amorphous pyroxene & 2.0 & $12.35$ & $10.88$ & $3.86$ & $11.48$ & 9.64 & 3.38\\
Amorphous pyroxene & 5.0 & $0.00$ & $0.00$ & $0.00$ & $0.00$ & 0.00 & 0.00\\
\hline
Crystalline enstatite & 0.1 & $2.65$ & $2.54$ & $3.56$ & $3.01$ & 2.94 & 0.40\\
Crystalline enstatite & 2.0 & $0.00$ & $0.00$ & $0.00$ & $0.00$ & 0.00 & 0.00\\
Crystalline enstatite & 5.0 & $14.32$ & $20.68$ & $13.76$ & $20.71$ & 17.37 & 3.33\\
\hline
Amorphous silica & 0.1 & $0.66$ & $0.36$ & $0.19$ & $0.54$ & 0.44 & 0.18\\
Amorphous silica & 2.0 & $0.00$ & $0.00$ & $0.00$ & $0.00$ & 0.00 & 0.00\\
Amorphous silica & 5.0 & $0.00$ & $0.00$ & $0.00$ & $0.00$ & 0.00 & 0.00\\
\hline
\end{tabular}
\end{table*}

\FloatBarrier

\section{Slab model fits}

\begin{figure*}[h!]
\centering
\includegraphics[width=0.95\textwidth]{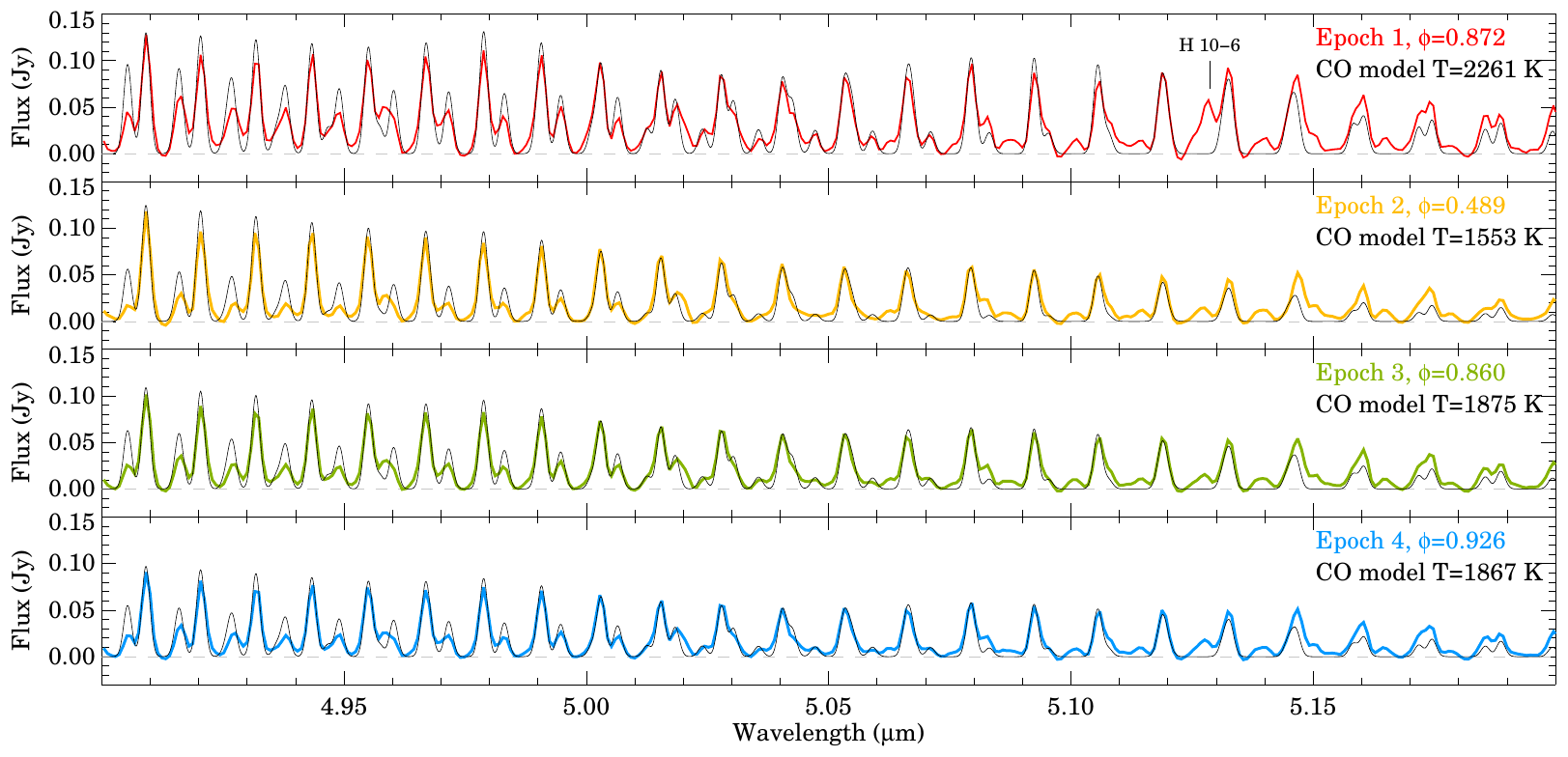}\\
\includegraphics[width=0.23\textwidth,page=1]{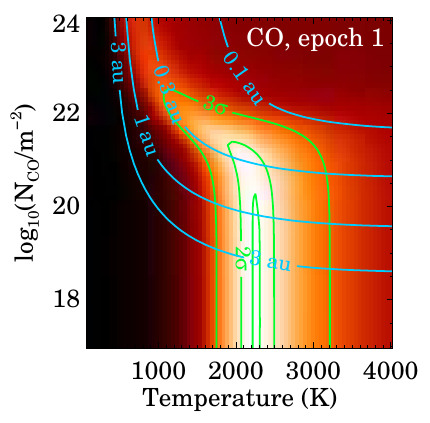}
\includegraphics[width=0.23\textwidth,page=2]{fit_single_temp_slab_co_chi2.pdf}
\includegraphics[width=0.23\textwidth,page=3]{fit_single_temp_slab_co_chi2.pdf}
\includegraphics[width=0.23\textwidth,page=4]{fit_single_temp_slab_co_chi2.pdf}
\caption{Top: observed spectra of DQ\,Tau (in color) with the best-fitting CO slab model fitted for the $4.9-5.1\,\mu$m range (black curves). Bottom: $\chi^2_{\rm min}/\chi^2$ surfaces with the 1$\sigma$, 2$\sigma$, 3$\sigma$ contours marked in green. The color scale goes from 0 (black) to 1 (white). The blue curves show the best-fitting slab radii for certain $N$-$T$ pairs assuming a circular emitting area.}\label{fig:co}
\end{figure*}

\begin{figure*}
\centering
\includegraphics[width=0.95\textwidth]{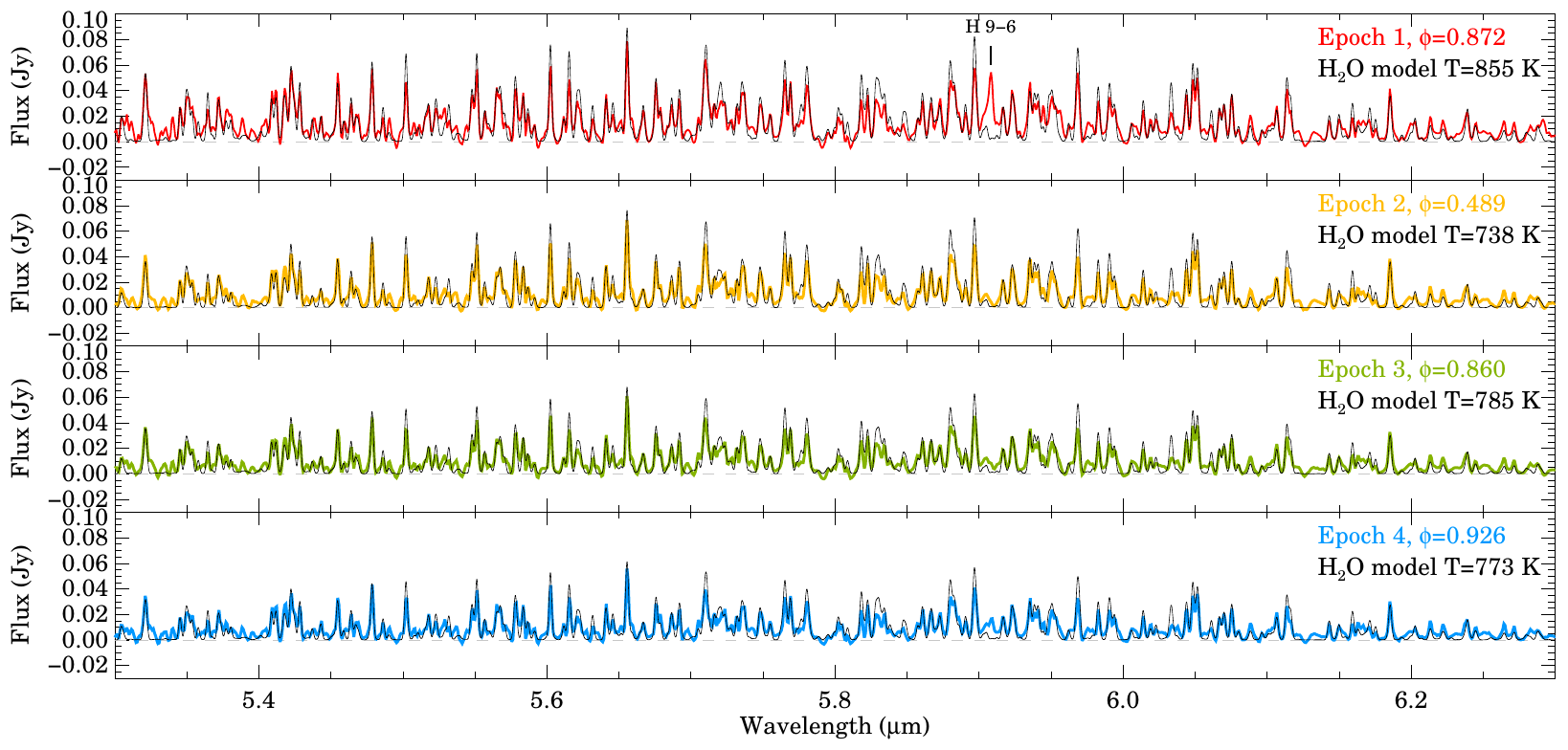}\\
\includegraphics[width=0.23\textwidth,page=1]{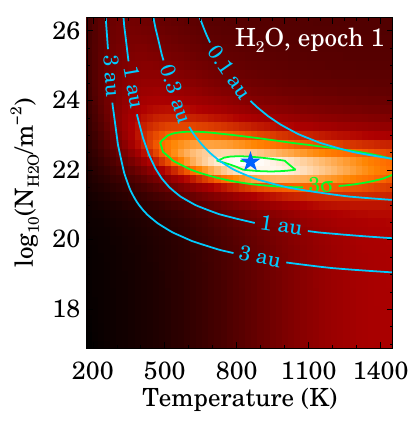}
\includegraphics[width=0.23\textwidth,page=2]{fit_single_temp_slab_h2o_5.3_6.3_chi2.pdf}
\includegraphics[width=0.23\textwidth,page=3]{fit_single_temp_slab_h2o_5.3_6.3_chi2.pdf}
\includegraphics[width=0.23\textwidth,page=4]{fit_single_temp_slab_h2o_5.3_6.3_chi2.pdf}
\caption{Top: observed spectra of DQ\,Tau (in color) with the best-fitting H$_2$O slab model fitted for the $5.3-6.3\,\mu$m wavelength range (black curves). Bottom: $\chi^2_{\rm min}/\chi^2$ surfaces with the 1$\sigma$, 2$\sigma$, 3$\sigma$ contours marked in green and the centroid marked with the blue asterisk. The color scale goes from 0 (black) to 1 (white). The blue curves show the best-fitting slab radii for certain $N$-$T$ pairs assuming a circular emitting area.}\label{fig:h2o}
\end{figure*}

\begin{figure*}
\centering
\includegraphics[width=0.92\textwidth]{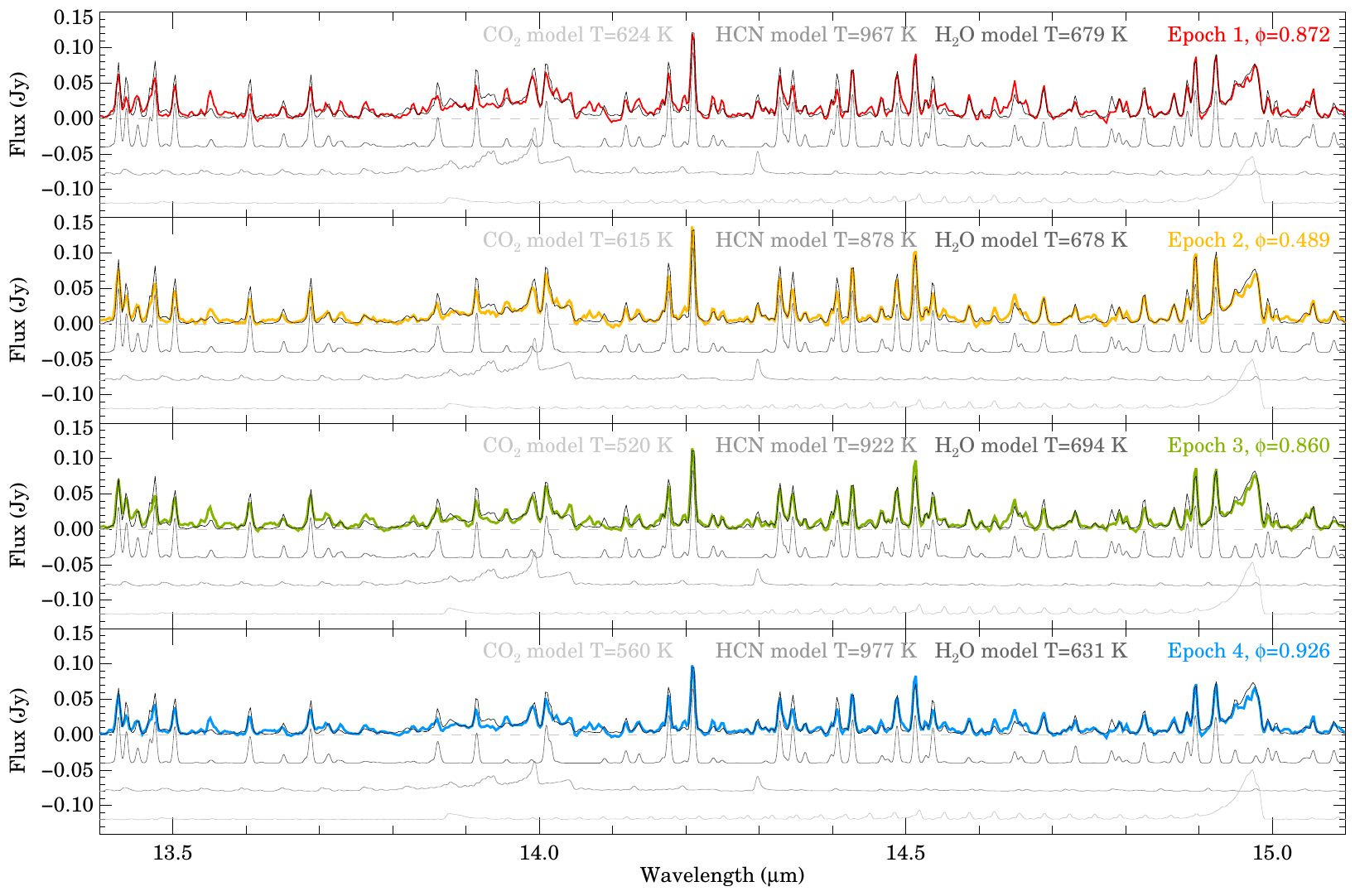}\\
\includegraphics[width=0.21\textwidth,page=1]{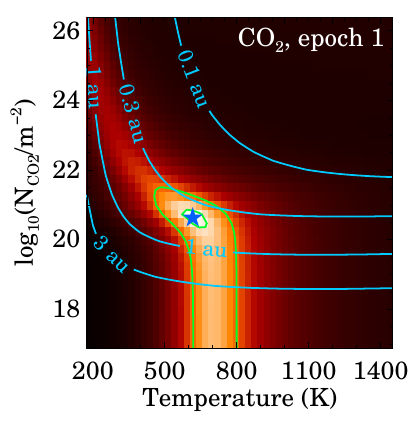}
\includegraphics[width=0.21\textwidth,page=2]{fit_single_temp_slab_co2_13.4_15.1_chi2.pdf}
\includegraphics[width=0.21\textwidth,page=3]{fit_single_temp_slab_co2_13.4_15.1_chi2.pdf}
\includegraphics[width=0.21\textwidth,page=4]{fit_single_temp_slab_co2_13.4_15.1_chi2.pdf}\\
\includegraphics[width=0.21\textwidth,page=1]{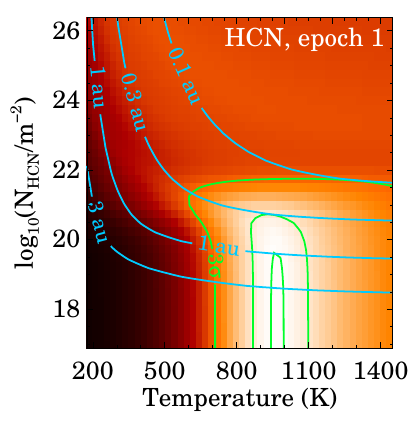}
\includegraphics[width=0.21\textwidth,page=2]{fit_single_temp_slab_hcn_13.4_15.1_chi2.pdf}
\includegraphics[width=0.21\textwidth,page=3]{fit_single_temp_slab_hcn_13.4_15.1_chi2.pdf}
\includegraphics[width=0.21\textwidth,page=4]{fit_single_temp_slab_hcn_13.4_15.1_chi2.pdf}\\
\includegraphics[width=0.21\textwidth,page=1]{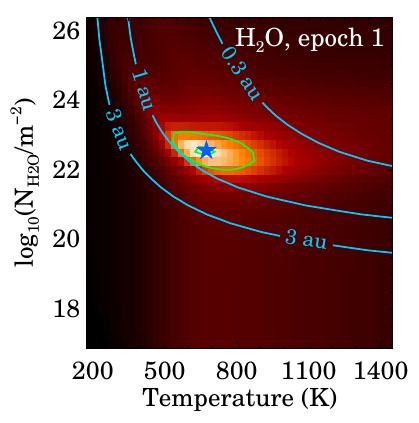}
\includegraphics[width=0.21\textwidth,page=2]{fit_single_temp_slab_h2o_13.5_15.0_chi2.pdf}
\includegraphics[width=0.21\textwidth,page=3]{fit_single_temp_slab_h2o_13.5_15.0_chi2.pdf}
\includegraphics[width=0.21\textwidth,page=4]{fit_single_temp_slab_h2o_13.5_15.0_chi2.pdf}
\caption{Top: observed spectra of DQ\,Tau (in color) with the best-fitting CO$_2$, H$_2$O, and HCN slab models fitted for the $13.4-15.1\,\mu$m wavelength range. For clarity, the models for the individual molecules (gray curves) were shifted along the y axis. The total model (in black) is the sum of the three individual models and is not shifted. Bottom: $\chi^2_{\rm min}/\chi^2$ surfaces with the 1$\sigma$, 2$\sigma$, 3$\sigma$ contours marked in green and the centroid marked with the blue asterisk. The color scale goes from 0 (black) to 1 (white). The blue curves show the best-fitting slab radii for certain $N$-$T$ pairs assuming a circular emitting area.}\label{fig:co2hcn}
\end{figure*}

\begin{figure*}
\centering
\includegraphics[width=0.95\textwidth]{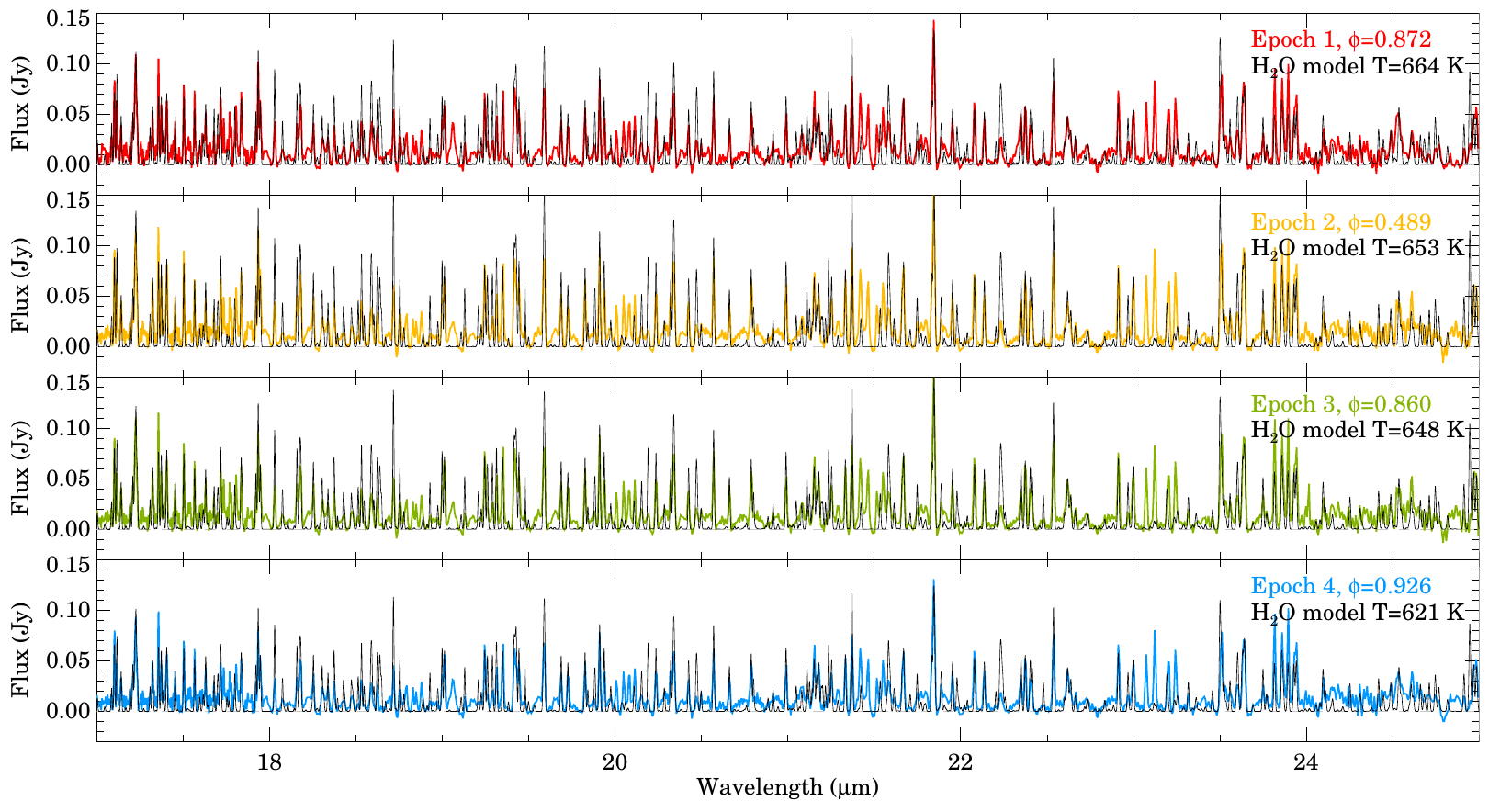}\\
\includegraphics[width=0.23\textwidth,page=1]{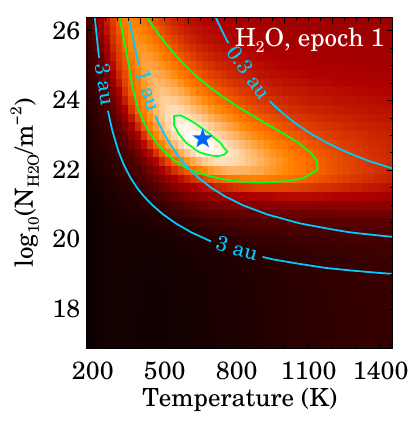}
\includegraphics[width=0.23\textwidth,page=2]{fit_single_temp_slab_h2o_17.0_25.0_chi2.pdf}
\includegraphics[width=0.23\textwidth,page=3]{fit_single_temp_slab_h2o_17.0_25.0_chi2.pdf}
\includegraphics[width=0.23\textwidth,page=4]{fit_single_temp_slab_h2o_17.0_25.0_chi2.pdf}
\caption{Top: observed spectra of DQ\,Tau (in color) with the best-fitting H$_2$O slab model fitted for the $17-25\,\mu$m wavelength range (black curves). Bottom: $\chi^2_{\rm min}/\chi^2$ surfaces with the 1$\sigma$, 2$\sigma$, 3$\sigma$ contours marked in green and the centroid marked with the blue asterisk. The color scale goes from 0 (black) to 1 (white). The blue curves show the best-fitting slab radii for certain $N$-$T$ pairs assuming a circular emitting area. Features not well accounted for by the model (typically groups of four lines) are from OH and the H\,I $8-7$ line at $19.06\,\mu$m.}\label{fig:h2o_long}
\end{figure*}

Figures~\ref{fig:co}, \ref{fig:h2o}, \ref{fig:co2hcn}, and \ref{fig:h2o_long} show the best-fitting slab models fitted to various wavelength ranges of the MIRI spectra.

\FloatBarrier

\end{appendix}

\end{document}